\documentclass[aps,showpacs,superscriptaddress,twocolumn]{revtex4-1}

\usepackage{graphicx}
\usepackage{color}
\usepackage{float}
\usepackage{subfigure}
\usepackage{braket}
\usepackage{amsmath}
\usepackage{amsthm}   
\usepackage{amsfonts}
\usepackage{amssymb}
\usepackage{epigraph}
\usepackage{fancyhdr}
\usepackage{wrapfig}
\usepackage{hyperref}
\usepackage{bm}

\begin{document}
\newcommand{\beq}{\begin{equation}}
\newcommand{\eeq}{\end{equation}}
\newcommand{\ben}{\begin{eqnarray}}
\newcommand{\een}{\end{eqnarray}}
\newcommand{\bea}{\begin{array}}
\newcommand{\eea}{\end{array}}
\newcommand{\om}{(\omega )}
\newcommand{\bef}{\begin{figure}}
\newcommand{\eef}{\end{figure}}
\newcommand{\leg}[1]{\caption{\protect\rm{\protect\footnotesize{#1}}}}
\newcommand{\ew}[1]{\langle{#1}\rangle}
\newcommand{\be}[1]{\mid\!{#1}\!\mid}
\newcommand{\no}{\nonumber}
\newcommand{\etal}{{\em et~al }}
\newcommand{\geff}{g_{\mbox{\it{\scriptsize{eff}}}}}
\newcommand{\da}[1]{{#1}^\dagger}
\newcommand{\cf}{{\it cf.\/}\ }
\newcommand{\ie}{{\it i.e.\/}\ } 
\setlength\abovedisplayskip{5pt}
\setlength\belowdisplayskip{5pt}

\title {Non-Hermitian Hamiltonian approach to quantum transport
in disordered networks with sinks: validity and effectiveness}

\author{Giulio~G. \surname{Giusteri}}
\affiliation{Dipartimento di Matematica e
Fisica and Interdisciplinary Laboratories for Advanced Materials Physics,
 Universit\`a Cattolica del Sacro Cuore, via Musei 41, I-25121 Brescia, Italy}
\affiliation{Istituto Nazionale di Fisica Nucleare,  Sezione di Pavia, via Bassi 6, I-27100, Pavia, Italy}
\affiliation{International Research Center on Mathematics \& Mechanics of Complex Systems, via XIX marzo 1, I-04012 Cisterna di Latina, Italy}
\author{Francesco \surname{Mattiotti}}
\affiliation{Dipartimento di Matematica e
Fisica and Interdisciplinary Laboratories for Advanced Materials Physics,
 Universit\`a Cattolica del Sacro Cuore, via Musei 41, I-25121 Brescia, Italy}
\author{G.~Luca \surname{Celardo}}
\affiliation{Dipartimento di Matematica e
Fisica and Interdisciplinary Laboratories for Advanced Materials Physics,
 Universit\`a Cattolica del Sacro Cuore, via Musei 41, I-25121 Brescia, Italy}
\affiliation{Istituto Nazionale di Fisica Nucleare,  Sezione di Pavia, via Bassi 6, I-27100, Pavia, Italy}

\begin{abstract}                
We investigate the validity of the non-Hermitian Hamiltonian approach
in describing quantum transport in disordered tight-binding networks
connected to external environments, acting as sinks. 
Usually, non-Hermitian terms are added, on a phenomenological basis, to such networks to summarize the effects of the coupling to the sinks.
Here we consider a paradigmatic model of open quantum network for which 
we derive a non-Hermitian effective model, discussing its limit of
validity by a comparison with the analysis of the full Hermitian
model.
Specifically, we consider
a ring of sites connected to a central one-dimensional
lead. The lead acts as a sink which absorbs the excitation initially present in
the ring. The coupling strength to the lead controls the opening of
the ring system. 
This model has been widely discussed
in literature in the context of light-harvesting systems.
We analyze the effectiveness of the non-Hermitian description both in absence and in presence 
of static disorder on the ring.
In both cases, the non-Hermitian model is
valid when the energy range determined by the eigenvalues of the ring Hamiltonian
is smaller than the energy band in the lead.
Under such condition, we show that results about the interplay of opening and disorder, previously
obtained within the non-Hermitian Hamiltonian approach, remain
valid when the full Hermitian model in presence of
disorder is considered.
The results of our analysis can be extended to generic networks with sinks, leading to 
the conclusion that the non-Hermitian approach is valid when
the energy dependence of the coupling to the external environments is sufficiently smooth in the energy range spanned by the eigenstates of the network.

\end{abstract}                                                               
                                                                            
\date{\today}

\maketitle

\section{Introduction}

Open quantum systems are nowadays at the center of many research fields 
in physics, ranging from quantum computing to 
transport in nano- and meso-scale solid state systems as well as biological aggregates.
In particular, charge/excitation transport in the quantum coherent regime
can be considered one of the central subjects in
modern solid state physics~\cite{Beenakker,Lee}. 
Transport properties depend strongly on the degree of
openness of the system.
In important applications, the effect of the opening is large, 
and cannot be treated perturbatively. The analysis of open quantum
systems beyond the perturbative regime is often difficult due to the
presence of infinitely many degrees of freedom.
Thus, a consistent way to take  
the effect of the opening into account for arbitrary coupling strength
between the system and the external world is highly desirable.

In a typical situation, we have a discrete quantum system coupled
to an external environment characterized by a continuum of states. 
Elimination of the continuum leads to an effective non-Hermitian 
Hamiltonian. This approach to open quantum
systems has been shown to be a very effective tool in dealing also with the strong coupling 
regime~\cite{MW,Zannals,rottertb,DittesR,Deffect2,ZeleReview}.
The non-Hermitian Hamiltonian approach offers several advantages: 
(i) it reduces an infinite dimensional problem to a finite dimensional
one; (ii) it allows to compute conductance and the whole
time-evolution of the relevant subsystem; (iii) the effects of
interference between discrete states and  the continuum, such as
Superradiance or Fano resonances can be easily analyzed~\cite{fgrluca}.  

Tight-binding networks are often considered in literature to model
quantum transport and decay, and their coupling with external environments,
acting as sinks, is taken into account by adding non-Hermitian terms to the Hamiltonian~\cite{kaplan,felix}.
Indeed, non-Hermitian models are more and more used to describe trapping or loss of excitation into transport channels of complex biological aggregates~\cite{deph,fassioli,srfmo}, but a proper justification of the employed non-Hermitian model is often overlooked.

Together with the coupling to a sink, such networks are usually
coupled to other environments, which induce different kinds of
disorder: static disorder (space-dependent) and dynamical disorder
(time-dependent). When disorder is added to the system to take into account the
effect of other environments, the strength of the coupling to the sink
is usually assumed to be unaffected by the disorder itself.

This assumption has been used both when dealing with dynamical disorder~\cite{mukameldeph,lussardi} and with static disorder~\cite{fassioli,alberto,CelGiuBor14}. Specifically,
some of the authors of this paper have previously analyzed the interplay of opening and static disorder in paradigmatic models of quantum transport, such as one-dimensional and three-dimensional tight-binding models~\cite{alberto,CelGiuBor14}.
Within the framework of the non-Hermitian Hamiltonian approach, it was found 
a novel cooperative regime characterized by the presence of subradiant
hybrid states. 
Moreover, cooperative robustness to disorder 
has been shown~\cite{CelGiuBor14} to play an important role in the dynamics of quantum systems with sinks.
As a matter of fact, all of those results were obtained assuming the coupling to the sinks to be independent of the disorder strength, even if we expect this assumption to fail for large disorder.

In this paper we consider a tight-binding network composed by a
ring-like structure coupled to a semi-infinite lead (Fig.~\ref{rl}).
This model has been discussed in several publications in literature
due to its relevance to light-harvesting complexes and to proposals of
bio-engineered devices for photon sensing~\cite{cao,superabsorb,sarovarbio,fassioli,mukameldeph,mukamelspano}.
Here we derive a
non-Hermitian Hamiltonian able to describe the transport properties of
the model. By comparing the results of the full Hermitian model with
the results obtained with the non-Hermitian model,
we want to assess
the limit of validity for the use of a non-Hermitian Hamiltonian to
model the decay properties in
presence of a sink. We will analyze in detail  the case with no disorder, while
for the case in presence of disorder our main goal is to give a qualitative
discussion of the limit of validity of the non-Hermitian approach and
to ascertain its reliability in reproducing the physics of the full
Hermitian system, focusing on the existence of subradiant hybrid states and cooperative robustness to static disorder.

In Sec.~\ref{sec:nhh} we introduce the 
non-Hermitian Hamiltonian approach to open quantum systems;
in Sec.~\ref{sec:model} we present our Hermitian model and we derive the corresponding non-Hermitian Hamiltonian, showing, in Sec.~\ref{sec:ST}, the effects of Superradiance in such a system.
We then analyze the validity and effectiveness of the non-Hermitian
model in reproducing the dynamics of the Hermitian system, in both
absence (Sec.~\ref{sec:limnodis}) and presence (Sec.~\ref{sec:dis}) of
diagonal disorder. In  Sec.~\ref{sec:random}, we show how our
results generalize to generic networks with sinks.
A summary of the results and their implications for the modeling of quantum sinks is given in the concluding Section.

\section{Derivation of the non-Hermitian Hamiltonian}\label{sec:nhh}

In this Section we present a standard derivation of the non-Hermitian
effective Hamiltonian.  
Alternative derivations can also be found in Ref.s~\cite{Zannals,Messiah,ZeleReview}.

Let us consider a discrete quantum system $A$,  interacting with another
system $B$, which represents the environment.
We assume that the subspace $A$ is spanned by $N_A$ discrete states
$|i\rangle$, while the subsystem $B$ represents the environment with states $|c,E\rangle$, where $c=1,\ldots,M$ is a discrete quantum number, labeling the decay channels,
and $E$ is another discrete quantum number, representing the energy (we will take the continuum limit of this quantum number later). 

In order to derive the effective non-Hermitian Hamiltonian which
describes the intrinsic system $A$, 
let us consider the projectors, within the Hilbert space of the total system $A+B$, on the two subsystems: 
\begin{equation}\label{P}
P_A=\sum_{i=1}^{N_A} |i\rangle \langle i|, \quad P_B=\sum_{c=1}^M \sum_{E=1}^{N_B} |c,E\rangle \langle c,E| \,.
\end{equation}

Under the orthogonality conditions $\langle i|j \rangle=\delta_{i,j}$, $\langle c, E| c', E'
\rangle=\delta_{c,c'} \delta_{E-E'}$, $\langle
i|c,E\rangle=0$, we can rewrite the total Hamiltonian of the system as
\begin{equation}
H =H_0+V= 
\begin{pmatrix}
H_{AA} & 0\\
0 & H_{BB} 
\end{pmatrix}+
\begin{pmatrix}
0& H_{AB}\\
H_{BA} & 0 
\end{pmatrix}\,,
\label{E0}
\end{equation}
where 
\begin{equation}\label{eq:HAA}
H_{AA}=P_AHP_A\,,\quad H_{AB}=P_AHP_B\,,
\end{equation}
and similarly for the other terms.

We can now define the unperturbed propagator $G_0(x)=(x-H_0)^{-1}$
and the total propagator $G(x)=(x-H)^{-1}$,
 related by the Dyson equation
\[
G(x)=G_0(x) + G_0(x) V G(x)\,,
\]
which gives rise to the following coupled equations for $G_{AA}=P_AGP_A$ and $G_{BA}=P_BGP_A$: 
\[
\left\{
\begin{aligned}
G_{AA}&\mbox{}=G^0_{AA} + G^0_{AA} H_{AB} G_{BA}\,,\\
G_{BA}&\mbox{}=G^0_{BB} H_{BA} G_{AA}\,.
\end{aligned}
\right.
\]
By substitution we obtain
\[
G_{AA}= G^0_{AA} + G^0_{AA} H_{AB} G^0_{BB} H_{BA} G_{AA}
\]
and, taking into account that $G^0_{BB} =(x-H_{BB})^{-1}$, we have
\[
G_{AA}(x) = \frac{1}{x-H_{AA} - H_{AB} \frac{1}{x-H_{BB}} H_{BA}}\,.
\]
From this expression we obtain an effective Hamiltonian, 
which defines the propagator
over the subspace $A$ and takes the form
\begin{equation}
H_{\rm eff}(x)= H_{AA} + H_{AB} \frac{1}{x-H_{BB}} H_{BA}\,.
\label{Heff1}
\end{equation}

The effective Hamiltonian, Eq.~\eqref{Heff1}, can be written in
an explicit form taking into account the orthogonality conditions of the
states in the subsystems $A$ and $B$. 
Without loss of generality, we assume that the total Hamiltonian is diagonal on the
subsystem $B$:
$$
\langle c, E| {H} |c', E'\rangle= E \delta_{c,c'} \delta_{E-E'}
$$
Using the projectors, Eq.~\eqref{P}, we have
\[
\begin{aligned}
H_{\rm eff}(x)&\mbox{}= \sum_{i,j} |i\rangle \langle i| H |j\rangle \langle j| \\
&\mbox{}+\sum_{c,E} \sum_{i,j} |i\rangle \langle i| H|c,E \rangle
\frac{1}{x-E} \langle c,E| H |j\rangle \langle j|\,.
\end{aligned}
\]

Let us now define the transition amplitudes between the intrinsic states
and the states of the environment: 
\begin{equation}
A_i^c(E)= \langle i| H|c,E \rangle\,.
\label{A}
\end{equation}
Taking the continuum limit
\[
\sum_{c,E} \rightarrow \sum_c \int\rho(E)\,dE
\]  
and using the identity 
\[
\frac{1}{x-x_0}= {\rm Pv}\left( \frac{1}{x-x_0}\right) \pm i \pi \delta(x-x_0)\,,
\]
the non-Hermitian Hamiltonian can be written as
\begin{equation}\label{eq:HeffE0}
H^\pm_{\rm eff}(x)=H_{AA}+\Delta(x)\mp\frac{i}{2} Q(x)\,,
\end{equation}
where
\begin{equation}
\begin{aligned}
Q_{ij}(x)&\mbox{}= 2 \pi \sum_c \int  A_i^c(E) (A_j^c(E))^* \rho(E)\delta(x-E)\,dE\\
&\mbox{}=2 \pi \sum_c A_i^c(x) (A_j^c(x))^* \rho(x)
\end{aligned}
\label{Q}
\end{equation}
and 
\begin{equation}
\Delta_{ij}(x)= \sum_c {\rm Pv}\int \frac{A_i^c(E) (A_j^c(E))^*
  \rho(E)}{x-E}\,dE\,,
\label{Delta}
\end{equation}
with $\rho(E)$ the density of states in the environment $B$.
 
The ambiguity in the sign of the last term in Eq.~\eqref{eq:HeffE0}, producing two distinct forms of the effective Hamiltonian, comes from the fact that the propagator $G^0_{BB}$, which appears in Eq.~\eqref{Heff1}, can be associated with either the forward or the backward evolution: the minus sign gives the forward-in-time evolution, i.e.
\begin{equation}\label{eq:Ut0f}
\theta(t-t_0)\mathcal U(t,t_0)=-\frac{1}{2\pi i}\int_{-\infty}^{+\infty}\frac{\exp\left[-\frac{i}{\hbar}x(t-t_0)\right]}{x-H^+_{\rm eff}(x)}\,dx\,,
\end{equation}
while the plus sign gives the backward-in-time evolution, i.e.
\begin{equation}\label{eq:Ut0b}
\theta(t_0-t)\mathcal U(t,t_0)=\frac{1}{2\pi i}\int_{-\infty}^{+\infty}\frac{\exp\left[-\frac{i}{\hbar}x(t-t_0)\right]}{x-H^-_{\rm eff}(x)}\,dx\,.
\end{equation}
Note that $\mathcal U(t,t_0)$ is the projection through $P_A$ of the full evolution operator of the system $A+B$.
Thus, if the
initial state of the total system has components only on the intrinsic system $A$,
its evolution under the
operator  $\mathcal U(t,0)$ in Eq.~(\ref{eq:Ut}) gives the projection
of the wave function of the total system over the intrinsic system.

From now on we will use the notation $H_{\rm eff}(x)$ for $H^+_{\rm eff}(x)$, referring to $H_{\rm eff}(x)$ as the effective Hamiltonian. 
To actually compute the evolution of an initial state, it is convenient to make use of the ($x$-dependent) basis of eigenstates of $H_{\rm eff}(x)$. Since $H_{\rm eff}(x)$ is in general non-Hermitian, due to the presence of the decay operator $Q(x)$, its eigenvalues 
\[
\mathcal E_r(x)=E_r(x)-\frac{i}{2}\Gamma_r(x)\,,\quad r=1,\ldots,N_A\,,
\]
are complex, and it has left and right eigenstates 
\begin{equation*}
H_{\rm eff}(x)|r,x\rangle={\cal E}_{r}(x)|r,x\rangle, \quad \langle \tilde{r},x|
H_{\rm eff}(x)=\langle\tilde{r},x|{\cal E}_{r}(x)\,,
\end{equation*}
which are bi-orthogonal, i.e.~the identity operator is given by
\[
1=\sum_r |r,x\rangle \langle \tilde{r},x|\,.
\]
Note that, when $H_{\rm eff}(x)$ is \emph{symmetric}, $\langle \tilde{r},x|$ equals the \emph{transpose} of $|r,x\rangle$, and not its Hermitian conjugate, as it happens in the case of Hermitian Hamiltonian operators. 

We observe that the decomposition of the identity given above is correct only when the eigenstates of the Hamiltonian form a complete set. This will be always true in the systems considered in this paper, but it is not always the case for non-Hermitian Hamiltonians. Indeed, the breakdown of such a condition for parameter-dependent non-Hermitian operators defines the so-called exceptional points in parameter space, whose study is relevant to many physical systems~\cite{expoints}.

Assuming now $t>0$, the evolution operator on states of the intrinsic system $A$ can be written as
\begin{equation}\label{eq:Ut}
\mathcal U(t,0)=\frac{i}{2\pi}\sum_r\int_{-\infty}^{+\infty}
\frac{e^{-\frac{i}{\hbar}xt}|r,x\rangle \langle \tilde{r},x|}{x-E_r(x)+\frac{i}{2}\Gamma_r(x)}\,dx\,.
\end{equation}
Due to the coupling between the intrinsic system and the environment, the total probability for an initially intrinsic state to remain in $A$ may not be conserved in time, this is why the evolution operator $\mathcal U$ is, in general, non-unitary. This property can be already gathered form Eq.~\eqref{eq:Ut}, but it will become more evident in the next Section.

In the case $[H_{\rm eff}(x),H^-_{\rm eff}(x)]=0$ (which is always true in the case $N_A=1$) we can write the evolution operator in a particularly useful form.
Indeed,
we can define, for any $r=1,\ldots,N_A$,
\[
G_r^{\pm}(x)=\frac{|r,x\rangle \langle \tilde{r},x|}{x-E_r(x)\pm\frac{i}{2}\Gamma_r(x)}\,,
\]
and express the propagator in the form
\begin{equation}
\label{eq:Gmatrix}
\begin{aligned}
G(x)&\mbox{}=\sum_r\left(G_r^{+}(x)-G_r^{-}(x)\right)\\
&\mbox{}=\sum_r\frac{-i\Gamma_r(x)|r,x\rangle \langle \tilde{r},x|}{[x-E_r(x)]^2+\frac{1}{4}\Gamma_r(x)^2}\,,
\end{aligned}  
\end{equation}
so that the evolution operator on states of the intrinsic system $A$ reads
\begin{equation}\label{eq:Utpm}
\mathcal U(t,0)=\frac{1}{2\pi}\sum_r\int_{-\infty}^{+\infty}
\frac{e^{-\frac{i}{\hbar}xt}\Gamma_r(x)|r,x\rangle \langle \tilde{r},x|}{[x-E_r(x)]^2+\frac{1}{4}\Gamma_r(x)^2}\,dx\,.
\end{equation}
This form of the evolution operator will be used in the next Sections.

\subsection{Energy-independent non-Hermitian Hamiltonian}\label{sec:eindepH}

The effective
non-Hermitian Hamiltonian, Eq.s (\ref{eq:HeffE0},\ref{Q},\ref{Delta})
can be greatly simplified if the $x$-dependence can be neglected.
In the case of a single quantum level of energy $E_0$, the effective Hamiltonian becomes a complex number and we have
\[
H_{\rm eff}(x)=E_0+\Delta(x)-\frac{i}{2}Q(x)\,.
\]
In order to see under which conditions the $x$-dependence can be neglected we can analyze the expression for the evolution operator given in Eq.~\eqref{eq:Utpm}, where $E_1(x)=E_0+\Delta(x)$ and $\Gamma_1(x)=Q(x)$. If $\Gamma_1(x)$ and $\Delta(x)$ are smooth and slowly varying function around $x=E_0$, the propagator, Eq.~\eqref{eq:Gmatrix}, can be well approximated around $E_0$ by setting
\begin{equation}\label{eq:gdE0}
\begin{aligned}
\Delta(x)&\mbox{}=\Delta(E_0)\,,\\
\Gamma_1(x)&\mbox{}=\Gamma_1(E_0)\,.
\end{aligned}
\end{equation}
With this approximation the evolution operator becomes
\[
\mathcal U(t,0)\approx e^{-\frac{i}{\hbar}(E_0+\Delta(E_0))t}e^{-\frac{1}{2\hbar}\Gamma_1(E_0)t}\,.
\]
Clearly, the range of times in which the latter will be a good approximation for the actual evolution will depend on how well the propagator is approximated by a Lorentzian function even for energies distant from $E_0$. 
Note that the approximated propagator implies an exponential decay of the unstable quantum state with a decay width
\[
\Gamma_1(E_0)=2\pi\sum_c|A_c(E_0)|^2\rho(E_0)\,,
\]
see Eq.~\eqref{Q}, which corresponds to the Fermi Golden Rule.
Hence, the problem of the validity of the energy-independent effective Hamiltonian in the case of a single state is formally equivalent to the problem of the validity of the exponential decay given by the Fermi Golden Rule of an unstable quantum state~\cite{pascazio,peres,pastawski}.

When we have many levels coupled to the same continuum the evolution operator in a generic situation is given by Eq.~\eqref{eq:Ut}. The problem of obtaining an energy-independent non-Hermitian Hamiltonian able to describe such evolution is now more delicate since, in this case, we may have different energies associated with the different levels. This general problem will be treated in a subsequent publication. 

On the other side,  when  $\Delta_{ij}(x)$ and $Q_{ij}(x)$ are smooth and slowly varying functions of $x$ in the whole physically relevant energy range, determined by the eigenvalues of the Hamiltonian of the intrinsic system, 
we can completely neglect the dependence on the energy of the initial state.
Under these conditions
we can obtain an energy-independent effective Hamiltonian by setting 
\begin{equation}\label{eq:gdE0matrix}
\begin{aligned}
\Delta(x)_{ij}&\mbox{}=\Delta(E_0)_{ij}\,,\\
Q(x)_{ij}&\mbox{}=Q(E_0)_{ij}\,,
\end{aligned}
\end{equation}
where $E_0$ is any energy lying in the relevant range. 
We can thus treat also the case of many levels coupled to the same continuum with an
energy-independent non-Hermitian Hamiltonian which reads:
\begin{equation}
H_{\rm eff} =H_{\rm AA}+\Delta -\frac{i}{2}Q\,.
\label{eeh}
\end{equation}
This approximation will be used in Sec.~\ref{sec:dis} to analyze a system in presence of disorder. 

The energy-independent non-Hermitian Hamiltonian behaves
as an effective Hamiltonian and allows us to compute the time
evolution of the projection of the total wave function on the
intrinsic system, see Eq.~(\ref{eq:Ut}). 
Indeed, we can expand any initial state of the intrinsic
system, over the
eigenstates of the effective Hamiltonian and its time evolution can be
determined as follows:
\begin{equation}
|\psi (t)\rangle= e^{-i H_{\rm eff} t/ \hbar} |\psi (0)\rangle =\sum_r e^{-i {\cal E}_r t/\hbar} |r\rangle \langle
\tilde{r}|\psi (0)\rangle .
\label{psit}
\end{equation}

Note that, for the case of a single particle/excitation in the intrinsic system and zero temperature in the external environment, the energy-independent non-Hermitian Hamiltonian description is equivalent to the standard master equation in Lindblad form obtained under the Born--Markov--Secular approximation~\cite{fgrluca}. Nevertheless, the non-Hermitian Hamiltonian approach is computationally much more efficient because one has to integrate only $N_A$ equations while, with the master equation approach, one has $O(N_A^2)$ equations to deal with.

\subsection{Superradiance}\label{sec:superrad}

A very important phenomenon that can be easily analyzed by means of
the energy-independent effective Hamiltonian, see Eq.~(\ref{eeh}),  is Superradiance. It is the cooperative effect which produces a strong inhomogeneity in the decay rates of the states of the intrinsic subsystem: some states, named superradiant, display large decay rates,
while the decay of some other states is very slow, sometimes even negligible. We note that such an effect is also known as ``resonance trapping'' and is present in many physical systems such as nuclei, microwave resonators, and optical resonators (see e.g.~\cite{restrap}).

The roots of this effect lie in the interference due to the competition of different states of the intrinsic subsystem to decay in the same channel in the continuum.
Considering a generic situation, let us assume that we can obtain an $x$-independent form of the terms $Q$ and $\Delta$ of Eqs.~\eqref{Q} and \eqref{Delta}, respectively.
Necessarily, the effective $Q$ possesses a factorized structure, since it is derived by the tensor product of the (rectangular) transition matrices $A^c_i$. It thus can have only as many non-zero eigenvalues as the number $M$ of decay channels. 
In the energy-independent approximation, $\Delta$ usually displays the same factorized structure of $Q$, and thus $[\Delta,Q]=0$. 

We must now distinguish two situations:
\begin{enumerate}
\item When $[H_{AA},Q]=0$, the eigenvalues $\mathcal{E}_r$ of $H_{\rm eff}$ are 
given by the sum of those of $H_{AA}$ (Eq.~\eqref{eq:HAA}), $\Delta$, and $-(i/2)Q$, so that we can have at most $M$ 
non-vanishing decay widths, while $N_A-M$ eigenstates of $H_{\rm eff}$ are not decaying at all.

\item When $[H_{AA},Q]\neq 0$, we encounter an additional effect, named Superradiance transition.
Indeed, the relative energy scale of the opening term $\Delta-(i/2)Q$ with respect to that of the intrinsic Hamiltonian $H_{AA}$ becomes important in determining how the decay width is distributed among the eigenstates of $H_{\rm eff}$: when the opening is weak the eigenstates will be close to the eigenstates of $H_{AA}$ and all of them will have a similar decay width; on the other hand, when the opening is strong, the eigenstates will approach those of $\Delta-(i/2)Q$, and only $M$ of them will have a significant decay width. We then see a transition from a non-superradiant regime (weak opening) to a superradiant regime (strong opening). 
This transition is not present in Case 1 above, where we are in a superradiant regime for any opening strength. 

\end{enumerate}

One could also consider the case $[\Delta,Q]\neq 0$, but under this condition 
it is not possible to  predict the behavior of the system regarding Superradiance on general grounds, and we need to look at the eigenvalues of the specific $H_{\rm eff}$ at hand.

\section{The Model}\label{sec:model}

\begin{figure}[t]
\centering
\includegraphics[width=7.6cm]{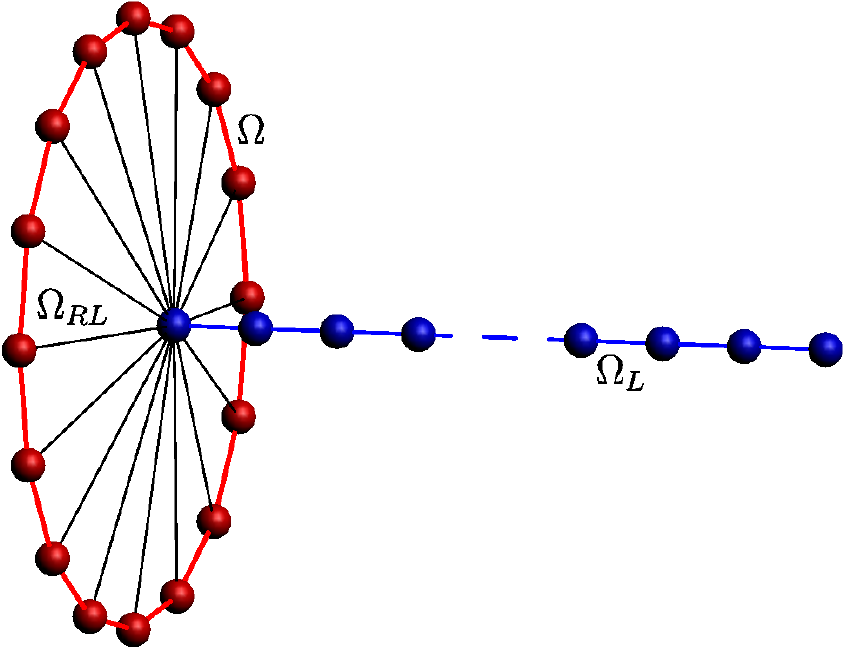}
\caption{Tight-binding model described by the Hermitian Hamiltonian $H$ given
in Eq.~\eqref{eq:HRL}: a ring of $N_R$ sites, connected with nearest-neighbor coupling $\Omega$, and a lead of $N_L$ sites, connected with nearest-neighbor coupling $\Omega_L$. The ring sites are equally coupled to the first lead site with tunneling amplitude $\Omega_{RL}$.
}
\label{rl}
\end{figure}

We consider here a simple model with $N_R$ two-level systems arranged in a ring structure
and coupled to a common decay channel, the sink, which we model with a one-dimensional lead. 

Such a ring-like structure has been considered in several papers 
\cite{cao,superabsorb,sarovarbio,fassioli,mukameldeph,mukamelspano} 
as a paradigmatic model
to describe different systems, such as 
molecular J-aggregates~\cite{Jaggr},
 bio-inspired devices for photon sensing~\cite{superabsorb}
and efficient light-harvesting systems~\cite{sarovarbio}.
In particular it has been often considered in the frame of exciton transport in
natural photosynthetic systems,
where chlorophyll molecules aggregate in ring-like structures around a reaction center, representing a central core absorber, where the excitation can be trapped~\cite{schulten1}. Chlorophyll molecules are able to absorb photons and can be modeled as two-level systems.
Under low light intensity, only one excitation is considered
and the molecular aggregate becomes
equivalent to a tight-binding model where one particle can hop from
site to site.

We first introduce a Hermitian model to describe the decay of excitation from the ring-like structure to the central core absorber represented by a lead, as described in Fig.~\ref{rl}. Note that also in Ref.~\cite{superabsorb} the central core absorber of the photon-sensing device was modeled by a lead.
Specifically, we consider a ring with $N_R$ sites, connected with nearest-neighbor coupling $\Omega$, described by the tight binding Hamiltonian
\begin{equation}
\label{HR}
H_R = \Omega \sum_{\langle r,r'\rangle}\left(|r\rangle \langle r'| + |r'\rangle \langle r|\right)\,, 
\end{equation}
where the sum runs over the pairs of neighboring sites.
In what follows we will measure energies in units of ${\rm cm}^{-1}$ and times in ${\rm ps}$. This choice, common in models for molecular aggregates, corresponds to setting $1/\hbar=0.06\,\pi\;{\rm cm/ps}$.

Each site of the ring is connected, through
the tunneling amplitude $\Omega_{RL}$, to the first site of a lead,
described by a linear chain of $N_L$ resonant sites with nearest-neighbor coupling $\Omega_{L}$. The Hamiltonian for the lead is
\begin{equation}
H_L= \Omega_L\sum_{j=1}^{N_L-1} \left( |\ell_j\rangle
\langle \ell_{j+1} | + |\ell_{j+1}\rangle
\langle \ell_j | \right)\,, 
\label{HL}
\end{equation}
and the interaction between the ring and the lead is described by
\begin{equation}
V_{RL}=\Omega_{RL} \sum_{r=1}^{N_R} \left( |r\rangle \langle \ell_1| + 
 |\ell_1 \rangle \langle r|\right) \,,
\label{VRL}
\end{equation}
so that the total Hamiltonian of the system, written on the site basis
\begin{equation}
\label{eq:sitebasis}
\left\{|r\rangle,|\ell_j\rangle,r=1,\ldots,N_R,j=1,\ldots,N_L\right\}\,,
\end{equation}
reads 
\begin{equation}
H= H_R + V_{RL} + H_{L}\,.
\label{eq:HRL}
\end{equation}

One can imagine that, when $N_L$ is large enough, the lead represents a
good sink, in that it absorbs most of the excitation present in the system. 

\subsection{The non-Hermitian Hamiltonian}\label{sec:modelA}

Since our main focus is on the decay of the excitation from the ring and not on its dynamics in the lead, we will now derive an effective Hamiltonian for
the subsystem formed by the ring, summarizing into non-Hermitian terms
the effects of the subsystem represented by the lead. In this derivation we will follow the procedure described in Sec.~\ref{sec:nhh}.

The eigenvalues of the lead Hamiltonian are given by
\begin{equation}
E_q=-2\Omega_L \cos k_q a\,,\quad q=1,\ldots,N_L\,,
\label{eq:leadev}
\end{equation}
where $a$ is the distance between adjacent sites and the wave number is
\[
k_q=\frac{\pi q}{a(N_L+1)}\,.
\]
The components on the site $|\ell_j\rangle$ of the lead eigenstates read
\begin{equation}\label{eq:leadstates}
\langle \ell_j|\psi_q\rangle=\sqrt{\frac{2}{N_L+1}} \sin{k_q j a}\,.
\end{equation}
We perform now the continuum limit taking 
$N_L\to \infty$. The discrete wave number $k_q$ becomes a continuous parameter ${k}\in (0,\pi/ a)$.
We obtain from Eq.~\eqref{eq:leadev}
\begin{equation}\label{eq:Ek}
E({k})=-2\Omega_L \cos{{k} a}
\end{equation}
and, recalling that
\[
\sin{{k} a}=\sqrt{1-\frac{E({k})^2}{4\Omega_L^2}}\,,
\]
we can derive the density of states $\rho(E)=d{k}/dE$ as
\[
\rho(E)=\frac{1}{2 \pi \Omega_L}
\frac{1}{\sqrt{1-(E/2\Omega_L)^2}}\,.
\]
Moreover, the eigenstates in the continuum are given by
\[
\langle {\ell_j}\vert\psi(E)\rangle=\sqrt{2}\sin{k}{ja}\,,
\]
with $E$ given by Eq.~\eqref{eq:Ek}, and their components on the first lead site (${j}= 1$) read
\[
\langle \ell_1\vert\psi(E)\rangle
=\sqrt{2}\sqrt{1-\frac{E^2}{4\Omega_L^2}}\,.
\]

We can now compute the matrix elements connecting the site $r$ of the
ring with the eigenstates of the lead with energy $E$: 
\begin{equation}
\begin{aligned}
A_r(E)&\mbox{}= \langle r|H|\ell_1 \rangle \langle \ell_1|\psi(E) \rangle\\
&\mbox{}=\Omega_{RL} \sqrt{2}
\sqrt{1-(E/2\Omega_L)^2}\,.
\end{aligned}
\end{equation}


Computing now the coupling terms
\begin{equation}\label{eq:AA}
A_r(E) A_{r'}(E)^*\rho(E)=\frac{\Omega_{RL}^2}{\pi\Omega_L}\sqrt{1-(E/2\Omega_L)^2}\,,
\end{equation}
we define the transition matrix $Q(x)$, see Eq.~\eqref{Q}, by
\begin{equation}
\label{eq:QNL}
Q_{rr'}(x)=\left\{
\begin{aligned}
\gamma\sqrt{1-\frac{x^2}{4\Omega_L^2}}\quad & \text{ for }x\in[-2\Omega_L,2\Omega_L]\,,\\
0\quad\qquad\quad & \text{ otherwise,}
\end{aligned}\right.
\end{equation}
where we introduced the opening strength
\begin{equation}\label{eq:gamma}
\gamma=\frac{2\Omega_{RL}^2}{\Omega_L}\,.
\end{equation}
By using Eq.~\eqref{eq:AA} we can also derive the expression for the matrix $\Delta(x)$, see Eq.~\eqref{Delta}, as
\begin{equation}
\label{eq:DeltaNL}
\Delta_{rr'}(x)=\frac{\gamma}{2\pi}{\rm Pv}\int_{-2\Omega_L}^{2\Omega_L}\frac{\sqrt{1-(E/2\Omega_L)^2}}{x-E}\,dE\,,
\end{equation}
thus obtaining the effective Hamiltonian
\begin{equation}
\label{eq:Hx}
H_{\rm eff}(x)=H_R+\Delta(x) -\frac{i}{2}Q(x)\,.
\end{equation}

Note that the matrix elements $Q_{rr'}(x)$ and $\Delta_{rr'}(x)$ do not depend on $r$ or $r'$. This fact implies that they commute and they both have only one non-zero eigenvalue. The state corresponding to that eigenvalue is
the fully symmetric state
\begin{equation}
|S\rangle =\frac{1}{\sqrt{N_R}} \sum_r |r\rangle\,,
\label{sr}
\end{equation}
which is also an eigenstate of $H_R$, corresponding to the maximum energy $2\Omega$.
The remaining $N_R-1$ eigenstates of $Q(x)$ and $\Delta(x)$ are degenerate and can always (i.e.~for any $x$) be chosen to match those of $H_R$ orthogonal to $|S\rangle$. For this reason, $Q(x)$ and $\Delta(x)$ commute with $H_R$.

The only non-zero eigenvalue of $Q(x)$ is given by
\begin{equation}\label{eq:GsrApp}
\Gamma_{\rm sr}(x)=
\left\{
\begin{aligned}
N_R\gamma\sqrt{1-\frac{x^2}{4\Omega_L^2}}\quad & \text{ for }x\in[-2\Omega_L,2\Omega_L]\,,\\
0\quad\qquad\quad & \text{ otherwise,}
\end{aligned}\right. 
\end{equation}
and the only non-zero eigenvalue of $\Delta(x)$ is
\begin{equation}\label{eq:DsrApp}
\Delta_{\rm sr}(x)=\gamma N_R {\rm Pv}\int_{-2\Omega_L}^{2\Omega_L} \frac{\sqrt{1-E^2/4\Omega_L^2}}{x-E}\,dE\,.
\end{equation}

From the foregoing facts, we obtain the important consequence that we can diagonalize the effective Hamiltonian $H_{\rm eff}(x)$
on the $x$-independent basis of eigenstates of $H_R$. The only eigenvalue of the intrinsic system (the ring) which is modified by the opening is
\[
\mathcal E_{\rm sr}=2\Omega+\Delta_{\rm sr}(x)-\frac{i}{2}\Gamma_{\rm sr}(x)\,,
\]
while the others are
\[
\mathcal E_r=E_r=2\Omega\cos\frac{2\pi r}{N_R}\,,\text{ for }r=1,\ldots,N_R-1\,,
\]
which coincide with the eigenvalues of $H_R$~\cite{CelGiuBor14}.

Remarkably, we are in the peculiar situation in which only one ring state ($|S\rangle$, Eq.~\eqref{sr}) is coupled to the lead,
and the number of relevant degrees of freedom, as far as decay properties are concerned, may look already dramatically reduced to $1$. Nevertheless, the dependency on $x$ of $\Gamma_{\rm sr}$ and $\Delta_{\rm sr}$, keeps the actual number of degrees of freedom infinite.

The time-evolution operator for the sole ring state, $|S\rangle$, which is coupled to the lead is given by
\begin{equation}
\label{eq:Usr}
\mathcal U_S(t,0)=\frac{1}{2\pi}\int_{-2\Omega_L}^{+2\Omega_L}
\frac{e^{-\frac{i}{\hbar}xt}\Gamma_{\rm sr}(x)}{[x-2\Omega-\Delta_{\rm sr}(x)]^2+\frac{1}{4}\Gamma_{\rm sr}(x)^2}\,dx\,,
\end{equation}
while the ring states which are orthogonal to $|S\rangle$ are effectively decoupled from the lead. For those states the opening term in the effective Hamiltonian vanishes, and it is trivially, but exactly, $x$-independent: those states will never decay.

\begin{figure}[t]
\centering
\includegraphics[width=8.5cm]{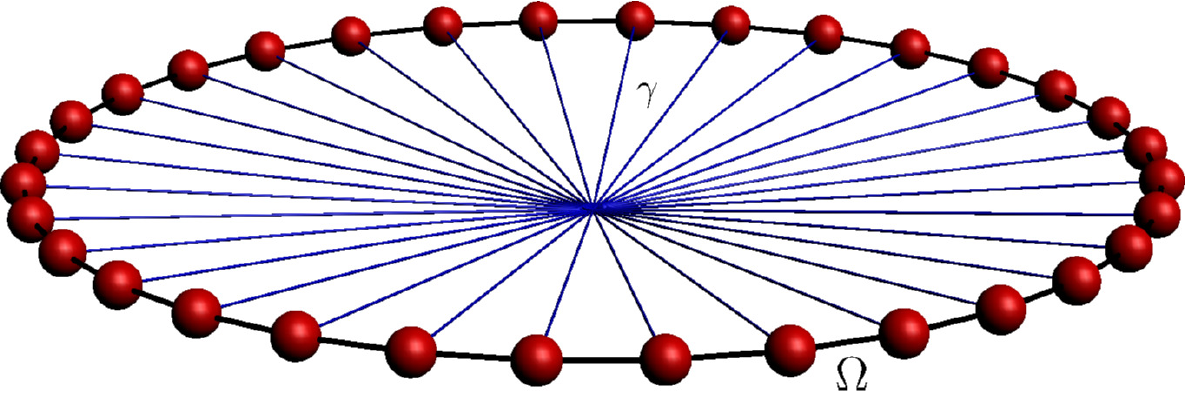}
\caption{Tight-binding model described by the effective Hamiltonian $H_{\rm eff}$ given
in Eq.~\eqref{eq:HeffR}: the $N_R$ resonant sites are coupled with nearest-neighbor tunneling amplitude $\Omega$. Moreover, they are equally open towards a common decay channel with opening strength $\gamma$.
}
\label{nonH}
\end{figure}

To complete the wanted dimensional reduction, we then need to derive an $x$-independent (or energy-independent) approximation of $H_{\rm eff}(x)$, Eq.~\eqref{eq:Hx}.

Now, if we let $\Omega_L\to\infty$ and $\Omega_{RL}\to\infty$ keeping $\gamma$ fixed (wide-band limit), we clearly obtain an exact energy-independence with 
\begin{equation}\label{eq:limDG}
\Delta_{\rm sr}(x)\to 0\quad\text{and}\quad \Gamma_{\rm sr}(x)\to \gamma N_R\,. 
\end{equation}
With those assumptions we get
\begin{equation}\label{eq:UteffR}
\mathcal U_S(t,0)=\exp\left(-\frac{2\Omega i}{\hbar} t-\frac{\gamma N_R}{2\hbar}t\right)\,,
\end{equation}
and the effective energy-independent non-Hermitian Hamiltonian describing the evolution of the intrinsic system, the ring, becomes
\begin{equation}
H_{\rm eff}=H_R -i\frac{\gamma}{2}\mathcal O\,,
\label{eq:HeffR}
\end{equation}
where $\mathcal O$ is a full matrix with all entries equal to $1$, and the components of $H_{\rm eff}$ on the ring-site basis read
\[
(H_{\rm eff})_{rr'}=(H_R)_{rr'} -i\frac{\gamma}{2}\,.
\]
Accordingly, the evolution operator on the whole intrinsic subspace is given by
\begin{equation}
\label{eq:UHeff}
\mathcal{U}(t,0)=e^{-iH_{\rm eff}t/\hbar}\,.  
\end{equation}

In summary, the effective non-Hermitian model, depicted in Fig.~\ref{nonH}, is given by an open ring of $N_R$ resonant sites equally coupled, with strength $\gamma$, to a common decay channel, in which the excitation can be lost.
We will analyze in Sec.~\ref{sec:limnodis} below the limit of validity of the energy-independent non-Hermitian Hamiltonian of Eq.~\eqref{eq:HeffR}.
Note that the non-Hermitian Hamiltonian just derived contains,
together with the Hamiltonian of the closed ring $H_R$, another term
$\mathcal O$, representing the decay
matrix. Since the matrix $\mathcal O$ is a full matrix, it represents a long-range hopping between the 
sites of the ring, mediated by the coupling of the sites of the ring
to the common decay channel in the lead.
This long range hopping will be relevant to understand the interplay of opening and disorder discussed in Sec.~\ref{sec:dis}.


\section{Superradiance in transport}\label{sec:ST}

The ring subsystem is, for any $\gamma \ne 0$,
in a superradiant regime, with a single superradiant state $|S\rangle$, Eq.~\eqref{sr}, absorbing all the decay width $\gamma N_R$, and $N_R-1$ subradiant states with vanishing decay widths.
Thus, our tight-binding model offers a paradigmatic realization of Case 1 of Sec.~\ref{sec:superrad}, showing that the symmetry of the coupling between each ring site and the sink is responsible for the effective perfect segregation of the decay widths.
Moreover, the Superradiance transition introduced in Case 2 of Sec.~\ref{sec:superrad} plays a fundamental role in determining the dynamics of our system when static disorder is added, and Sec.~\ref{sec:dis} is devoted to the analysis of such effect in both the Hermitian and the non-Hermitian models introduced above.

\begin{figure}[t]
\centering
\includegraphics[width=8.cm]{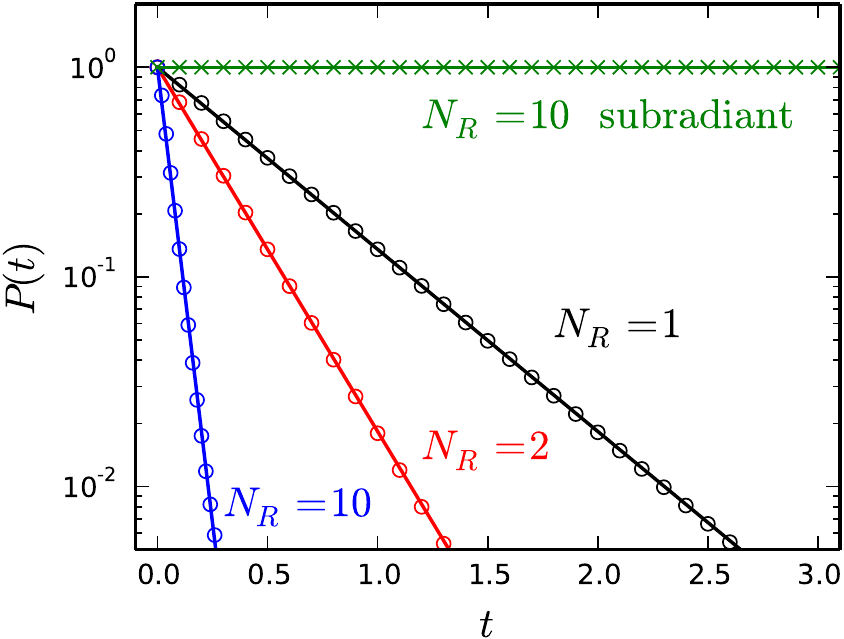}
\caption{(Color online) Survival probability $P(t)$ {\it vs} time $t$. 
Results obtained with the Hermitian model, Eq.~\eqref{eq:HRL}, (symbols) are compared with the results obtained with the non-Hermitian model, Eq.~\eqref{eq:HeffR}, (curves) for different values of $N_R$. 
Circles represents data obtained starting from the fully symmetric superradiant state $|S\rangle$ of Eq.~\eqref{sr}, while crosses refer to the antisymmetric state of Eq.~\eqref{eq:asym}, which
is subradiant. Values of the parameters are $\Omega=1$, $\Omega_{RL}=10$, $\Omega_L=100$, $\gamma=2$, and $ N_L=1000$.
}
\label{rl2}
\end{figure} 

Here we consider the effects of Superradiance on the decay of states which are initially excited on the ring, showing that the non-Hermitian model correctly reproduces the dynamics of the full Hermitian system. Given an initial ring state $|\psi_0\rangle$, we consider the survival probability 
\begin{equation}
\label{pstay}
P(t) = \sum_{r=1}^{N_R} | \langle r |\psi (t) \rangle |^2\,,
\end{equation}
computing the time evolution in both the Hermitian model, Eq.~\eqref{eq:HRL},
\begin{equation}
|\psi^H (t) \rangle=e^{-iHt/\hbar}|\psi_0 \rangle\,,
\end{equation}
and the non-Hermitian one, Eq.~\eqref{eq:HeffR},
\begin{equation}
|\psi^{H_{\rm eff}} (t) \rangle=e^{-iH_{\rm eff}t/\hbar}|\psi_0 \rangle\,.
\end{equation}

In Fig.~\ref{rl2} we compare the $P(t)$ obtained with the two models for the superradiant initial state of Eq.~\eqref{sr}, varying the system size as $N_R=1,2,10$ (black, red, and blue data, respectively). The agreement between the Hermitian model (circles) and the non-Hermitian one (curves) is excellent. Moreover, the decay width, given by
\[
\frac{N_R 2\Omega_{RL}^2}{\Omega_L}=N_R\gamma\,,
\]
increases well above the single-site decay rate $\gamma$ as $N_R$ is increased, signaling the presence of cooperative effects in the dynamics.

The same excellent agreement between the Hermitian model (green crosses) and the non-Hermitian one (green curve) is found in the $P(t)$ computed for the initial state
\begin{equation}
\label{eq:asym}
| AS \rangle=\frac{1}{\sqrt{N_R}}\sum_r (-1)^r |r\rangle\,,
\end{equation}
which is remarkably different from the one computed for $| \psi_0
\rangle=|S \rangle$. Indeed, as anticipated above by analyzing the
non-Hermitian model, we are in a superradiant regime. The state $| AS
\rangle$ is a subradiant eigenstate of $H_{\rm eff}$ with vanishing
decay width, and then its survival probability is constantly equal to
$1$. This supprression of decay, due to interference effects, is somehow surprising if one consider
that all the sites are coupled to a semi-infinite lead and
nevertheless the excitation never leaves the system. 

It is important to stress that the super/subradiant dynamics, predicted on the basis of the reduced non-Hermitian system, faithfully reproduces the Hermitian evolution, at least up to the times shown in the figure. For larger times or for very short times the behavior of the two models will depart from each other, due to the fact that, in our simulations, both $\Omega_L$ and $N_L$ are finite. In Sec.~\ref{sec:limnodis} we will analyze in detail the critical times up to which the agreement persists.

\section{Limit of validity of the non-Hermitian model: non-exponential decay}\label{sec:limnodis}

The non-Hermitian Hamiltonian, Eq.~(\ref{eq:HeffR}), constitutes a great simplification of the full Hermitian problem, since it eliminates the infinite number of degrees of
freedom of the lead. As it was shown in the previous Sections, the
non-Hermitian Hamiltonian becomes exact for infinite length and
infinite energy band in the lead. 

In this Section we want to clarify the effect of $\Omega_L$ and $N_L$
being finite on the validity of the non-Hermitian approach. 
We will restrict our attention to the survival
probability $P(t)$ computed for the superradiant initial state
$|S\rangle$ of Eq.~\eqref{sr}, for which 
the non-Hermitian Hamiltonian predicts an exponential decay with a
decay width given by $N_R \gamma$.
For all the other initial states, orthogonal to the state $|S\rangle$,
the non-Hermitian Hamiltonian  predicts that they  are subradiant and do not decay at all.  
Since we have only one level, $|S\rangle$, coupled to the lead, the problem of the validity of the non-Hermitian Hamiltonian approach is 
formally equivalent to the validity of 
the exponential decay, given by 
the Fermi Golden Rule, of
the survival probability of a single unstable quantum state coupled with a
continuum of states~\cite{pascazio,peres,pastawski,ZeleNew}. Also in our case we will show that 
the exponential behavior is typically valid for intermediate times,
while for both short and long times the decay is not exponential.

\subsection{Finite-size effects}

For finite lead length the decay of the excitation from the ring will not
be irreversible, since the excitation can bounce
back at the end of the lead, inducing a revival of the
survival probability.
Thus, we can expect deviations of the Hermitian evolution from the
exponential decay due to the reflection of the wave-packet at the end of the lead. This bouncing effect is always present in any finite-size sink and it is known in literature as mesoscopic echo~\cite{pastawski}.

We can analytically estimate the bouncing time $t_{B}$ assuming that the excitation,
after leaving the ring, goes through the lead with a certain group velocity $v_g$, and bounces back once reached the end of the lead.

We can describe the eigenstates of the lead as plane waves (see Eq.~\eqref{eq:leadstates})
\begin{equation}
\langle \ell_j| \psi_q\rangle \propto \sin k_q j a\,, \qquad j=1,\dots,N_L\,,
\end{equation}
with wave numbers
\begin{equation}
 k_q = \frac{\pi q}{a(N_L+1)} \, , \qquad k_q \in (0,\pi/a) \, .
\end{equation}
From this point of view, a superposition of lead eigenstates forms a wave-packet. The group velocity of this wave-packet is given by
\begin{equation}
 v_g = \left. \frac{\partial \omega (k)}{\partial k} \right|_{\overline{k}} =\frac{1}{\hbar} \left. \frac{\partial E(k)}{\partial k}
 \right|_{\overline{k}}\,,
\end{equation}
where $\overline{k}$ is the mean wave number of the waves that form the wave-packet.
Using the wave numbers of the eigenstates defined above, we can write the energies of the lead, Eq.~\eqref{eq:leadev}, as
\begin{equation}
 E(k_q) = -2 \Omega_L \cos k_qa \, ,
\end{equation}
so that the group velocity becomes
\begin{equation}\label{eq:vg}
 v_g =\frac{ 2 \Omega_L a}{\hbar} \sin \overline{k} a\, .
\end{equation}
The excitation has to go through the entire lead twice
before reaching again the starting point. From Eq.~\eqref{eq:vg}, we see that the maximum velocity of a wave-packet is $v_g=2\Omega_La/\hbar$. Hence, we can expect the agreement between the Hermitian evolution and the non-Hermitian one to persist up to the time
\begin{equation}\label{eq:tbounce}
t_B = \frac{2 a N_L}{v_g} =\frac{\hbar N_L}{\Omega_L}\, .
\end{equation}

In order to check our estimate for $t_B$, 
in Fig.~\ref{f22} we compare the superradiant decay $\exp(-N_R\gamma
t)$, produced by the non-Hermitian model (green curve), with the
Hermitian evolution computed for different values of the
lead size $N_L$. As $N_L$ increases, the agreement between the
Hermitian and the non-Hermitian evolution persists up to a critical time after which we have deviations from the exponential decay and a revival of the survival probability.
For small values of $N_L$ the agreement time increases linearly
with $N_L$  and it is well estimated by the values of $t_B$, see
vertical arrows.
On the other side, for larger values of $N_L$, the agreement time becomes independent of the length of the lead.
This suggests that a different effect, see discussion below, causes the departure of $P(t)$ from the exponential decay.

\subsection{Finite-bandwidth effects: $\Omega\ll\Omega_L$} 

\begin{figure}[t]
\centering
\includegraphics[width=8.5cm]{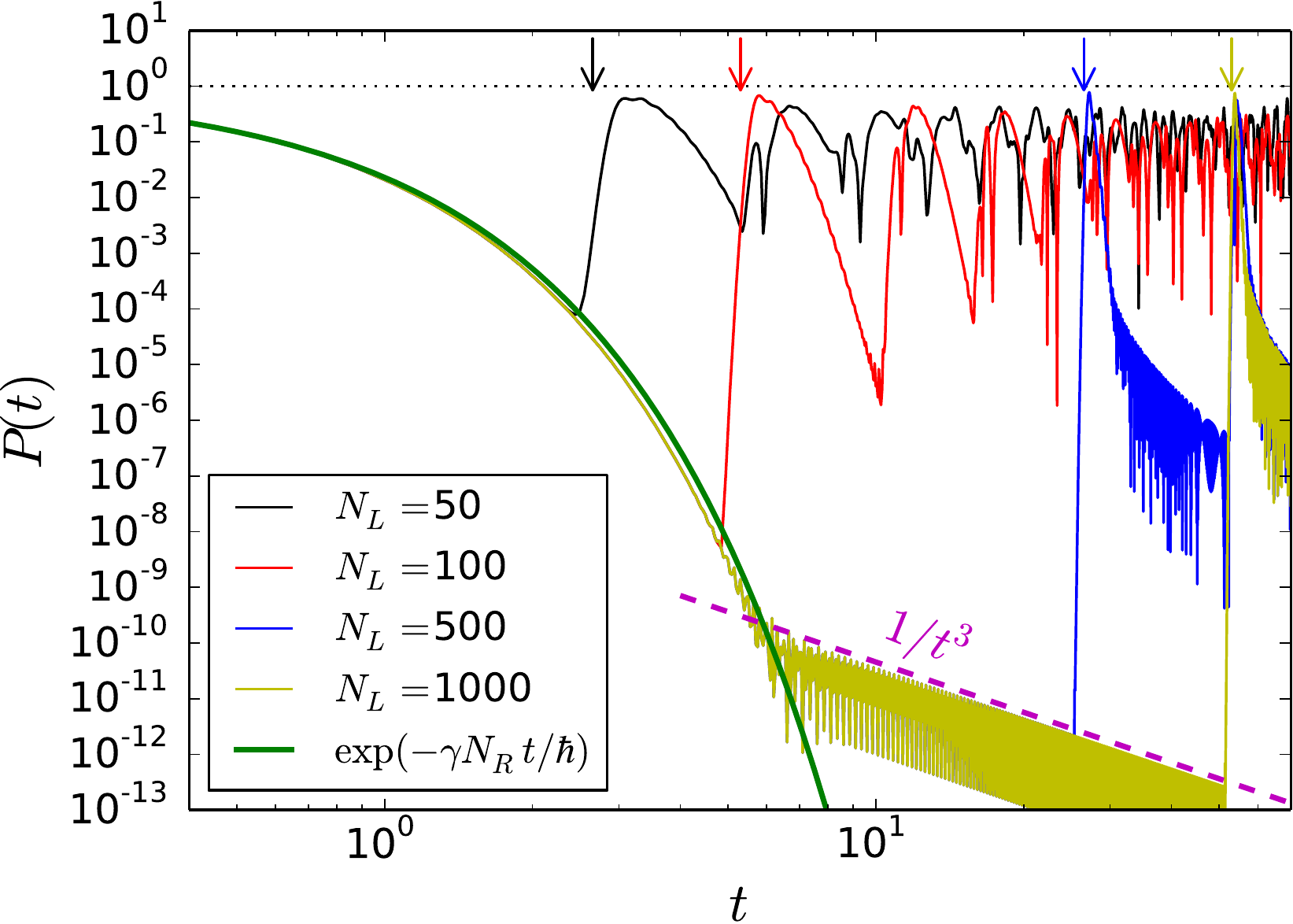}
\caption{(Color online)
The survival probability $P(t)$ computed starting from the superradiant state of Eq.~\eqref{sr} is plotted  {\it vs} time. 
The analytic decay $\exp(-N_R\gamma t)$ (green curve), obtained from the non-Hermitian model with $\gamma$ given by Eq.~\eqref{eq:HeffR}, is compared with the data obtained from the Hermitian model with $N_R = 10$,
$\Omega = 1$, $\Omega_{RL} = 10$, and $\Omega_L = 100$ for different values of $N_L$. 
Vertical arrows mark the values of $t_B$ given by Eq.~\eqref{eq:tbounce} for the corresponding values of $N_L$.
}
\label{f22}
\end{figure}

From Fig.~\ref{f22} we notice that, when we increase $N_L$ above a certain value, the
agreement time does not improve, even if the bouncing time increases.
Indeed, the large-size regime is characterized by an $N_L$-independent
agreement time, marking the transition from the superradiant decay to
a much slower one. The origin of this brake in the decay is very
general and it is due to the presence of a finite energy band in the
lead, whose bandwidth equals $4\Omega_L$, see discussion in Ref.~\cite{pascazio}.

Indeed from Eq.~(\ref{eq:Usr}) we see that the time evolution of the
superradiant state is given by the Fourier transform of the function
\begin{equation}
{\cal L} (x)= \frac{1}{\pi} \frac{\Gamma_{\rm sr}(x)/2}{[x-2\Omega-\Delta_{\rm sr}(x)]^2+\frac{1}{4}\Gamma_{\rm sr}(x)^2}\,.
\label{lor}
\end{equation}

In the limit  $\Omega_L\to\infty$, $\Gamma_{\rm sr}$ and $\Delta_{\rm sr}$
do not depend on $x$ (see discussion in Sec.~\ref{sec:model}).
Moreover, the limits of integration in Eq.~\eqref{eq:Usr} go to infinity. Thus, in this limit,
the time evolution of the superradiant state is the Fourier transform
of a Lorentzian function, which gives an exponential decay. On the
other side, for finite bandwidth in the lead, we can expect deviations
from the exponential decay due to two reasons: (i) the Lorentzian function
is now distorted due the fact that both $\Gamma_{\rm sr}$ and $\Delta_{\rm sr}$
depend on $x$ and (ii) the limits of integration do not go to infinity
anymore.

In this Subsection we will consider the situation in which the energy range of the ring is much smaller than the energy band in the lead, so that
the transition amplitude
$A_r(E)$ and the density of states $\rho(E)$, see Eq.~\eqref{eq:AA},
are very smooth and slowly varying function of the energy in the whole energy range of the ring. 
We are thus allowed to set the width $\Gamma_{\rm sr}(x)=\Gamma_{\rm sr}(0)=N_R\gamma$, see Eq.~\eqref{eq:GsrApp}, and $\Delta_{\rm sr}(x)=\Delta_{\rm sr}(0)=0$, see Eq.~\eqref{eq:DsrApp}.
For this reason our results will not depend on the energy $2\Omega$ of the initial state $|S\rangle$ and we can use the approximation $\Omega\approx 0$.
This regime is characterized by the following conditions:
\begin{equation}
\label{eq:condE0}
\Omega\ll\Omega_L\,,  \qquad \Gamma_{\rm sr}=N_R \gamma \ll 4 \Omega_L\,.
\end{equation}

In Appendix A, using the conditions in Eq.~\eqref{eq:condE0}, we show
that the function  $\mathcal L(x)$ is well approximated by a Lorentzian function.
In such a case, the main deviations from the exponential decay are due solely to the truncation of the function $\mathcal L(x)$ outside the energy band of the lead. 

Specifically, in Appendix A, we show that the decay of $P(t)$ is exponential between two time-scales $t_0$ and $t_S$ and we have:
\begin{equation}\label{eq:Pt3}
P(t)\approx\left\{
\begin{array}{cl}
 1-\frac{N_R\Omega_{RL}^2}{\hbar^2}t^2 & \quad\text{for }t<t_0\,,\\
&\\
 e^{-\Gamma_{\rm sr} t/\hbar}& \quad\text{for }t_0<t<t_S\,,\\
& \\
\mathrm{const.}/t^3 & \quad\text{for }t>t_S\,.
\end{array}\right.
\end{equation}

From Eq.~\eqref{eq:Pt3} we see that the decay is initially quadratic in time, as predicted by perturbation theory, then it is exponential with a decay width $\Gamma_{\rm sr}=N_R \gamma$ given by the Fermi Golden Rule, and eventually it decays as a power law.

The transition from quadratic to exponential decay occurs at a time $t_0$ given by
\begin{equation}\label{eq:t0noapp}
t_0=\frac{\hbar}{2\Omega_L}\,,
\end{equation}
which has been derived in Appendix A and, with a more heuristic approach, in Appendix B.

\begin{figure}[t]
\centering
\includegraphics[width=8.5cm]{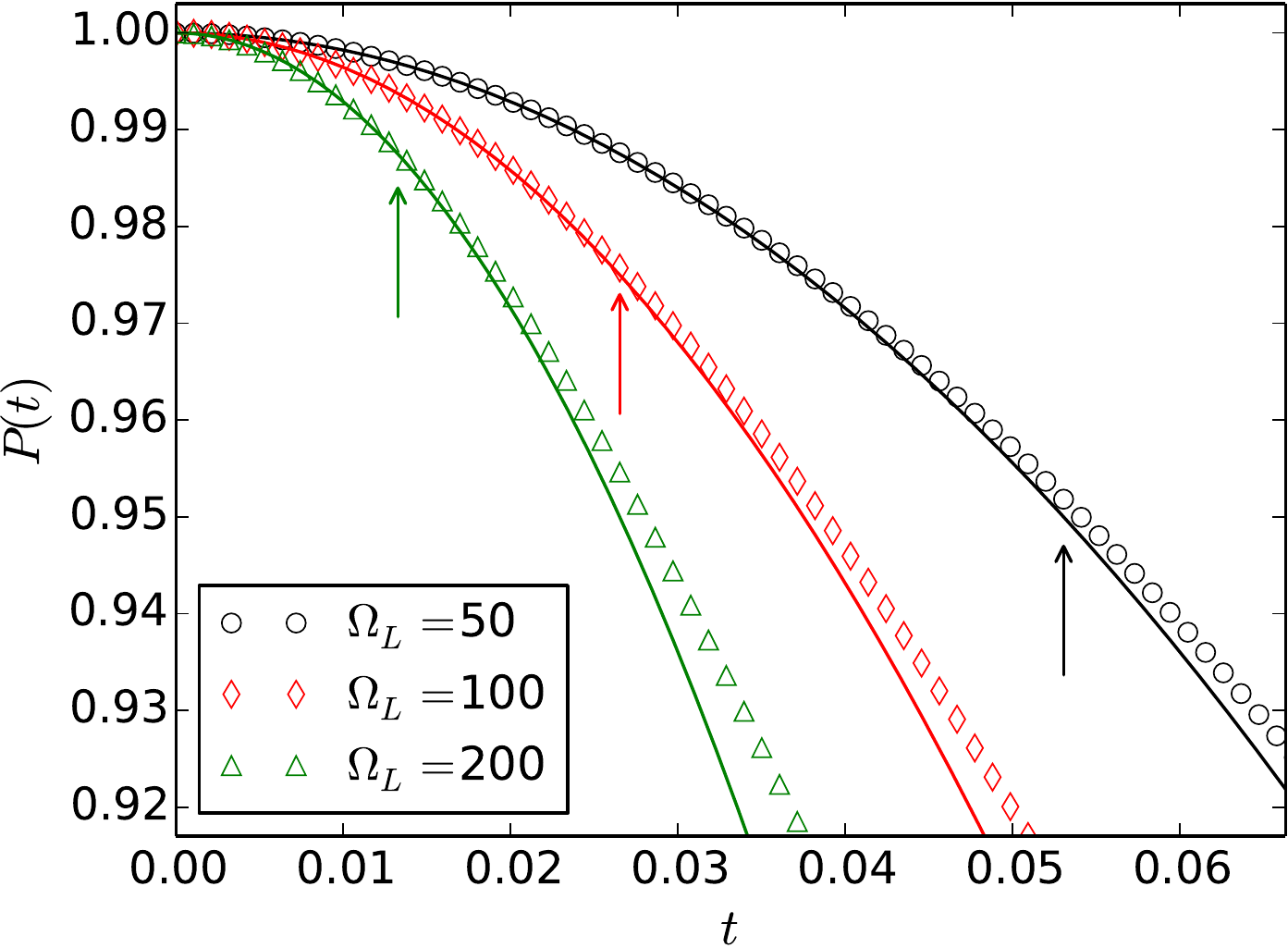}
\caption{(Color online)
Evolution of the survival probability $P(t)$, computed starting form the state $|S\rangle$, Eq.~\eqref{sr}, is plotted {\it vs} time $t$
 for different values of $\Omega_L$. The evolution of the
Hermitian model (symbols) is well approximated by the parabolic decay
(full curves) of Eq.~\eqref{eq:Pt3} for times shorter than $t_0=\hbar/2\Omega_L$ (vertical arrows).
Parameters are $\Omega=1$, $N_R=10$, $N_L=100$, and $\Omega_{RL}=\sqrt{\Omega_L}$ so that we keep the opening $\gamma=2$ fixed.
}
\label{fig:shortt}
\end{figure}

The quadratic initial decay given by the full Hermitian model is shown in Fig.~\ref{fig:shortt} (symbols) for diffrent values of $\Omega_L$. In the same figure, the short-time anlytic estimate (curves) given in Eq.~\eqref{eq:Pt3} is shown to be 
a good estimate of the initial behavior of the Hermitian system up to the time $t_0$, marked with arrows in Fig.~\ref{fig:shortt}.

\begin{figure}[t]
\centering
\includegraphics[width=8.5cm]{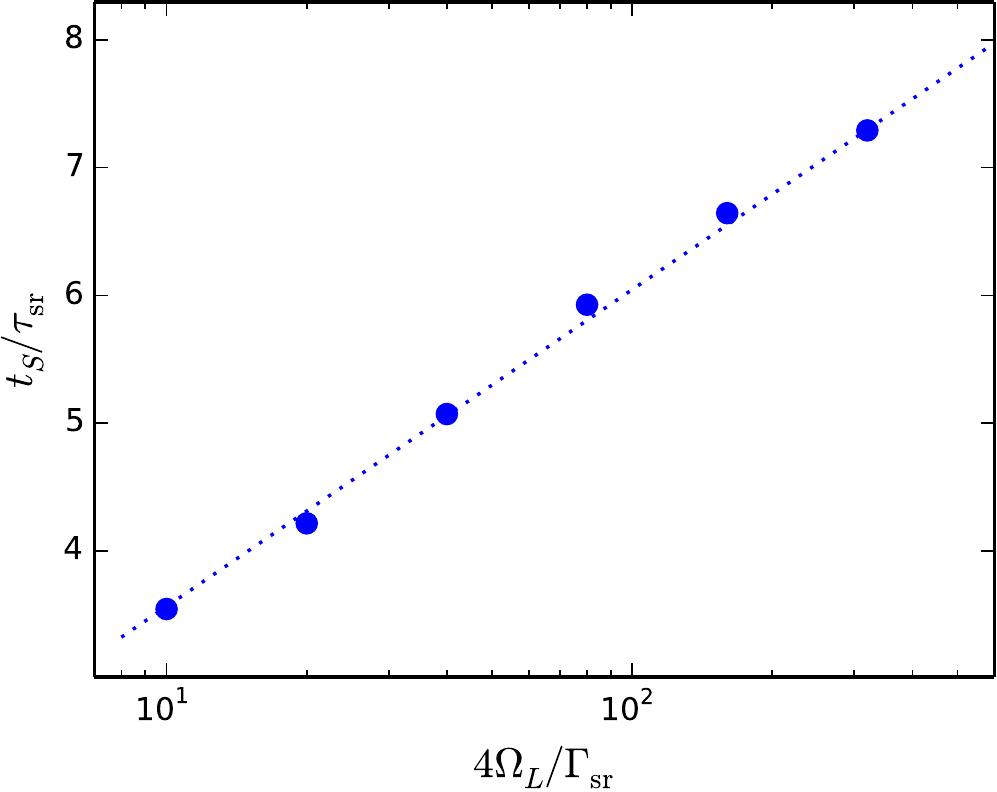}
\caption{(Color online) The ratio between the time $t_S$ at which the decay of the
  survival probability $P(t)$, computed with the Hermitian evolution
  of $|S\rangle$, departs from the exponential decay predicted by the
  non-Hermitian model and the characteristic decay time $\tau_{\rm sr}=\hbar/\Gamma_{\rm sr}$ is plotted against the ratio $4\Omega_L/\Gamma_{\rm sr}$. Data are obtained keeping $\Omega_L=\Omega_{RL}^2$ and $N_R=10$, so that the decay rate $\gamma=2$ is fixed.
We considered a   fixed value of $\Omega=1$ (circles).
The logarithmic scaling predicted in Eq.~\eqref{eq:t2} is apparent and it has been highlighted by means of the dotted curve.
}
\label{f24}
\end{figure}

The power-law decay $P(t)\propto t^{-3}$ for $t>t_S$
is in agreement with numerical results, see dashed line in Fig.~\ref{f22}.
The critical time $t_S$ for which we have the transition from the exponential
decay to the power-law decay has also been derived in Appendix A and we have: 
\begin{equation}
\label{eq:t2}
t_S\propto \frac{\hbar}{\Gamma_{\rm sr}}\ln\frac{4\Omega_L}{\Gamma_{\rm sr}}\,.
\end{equation}
For the exponential decay to be a good approximation on a significant time range we need $t_S$ to be several times the mean life-time $\tau_{\rm sr}=\hbar/\Gamma_{\rm sr}$. This can be achieved only if the logarithmic term in Eq.~\eqref{eq:t2} is large enough.
In Fig.~\ref{f24} the logarithmic dependence of $t_S$ on
$\Omega_L$ is shown to agree with numerical results.

We also observe that both the finite-size and finite-bandwidth
effects disappear in the thermodynamic limit ($N_L\to\infty$,
$\Omega_L\to\infty$, $\rho(E)=1/2\pi$) considered in
Sec.~\ref{sec:model} during the derivation of $H_{\rm eff}$, since
both $t_B$ and $t_S$ grow to infinity, while $t_0$ goes to zero.


\subsection{Finite-bandwidth effects: $\Omega\sim\Omega_L$}\label{sec:strong}

\begin{figure}[t]
\centering
\includegraphics[width=8.5cm]{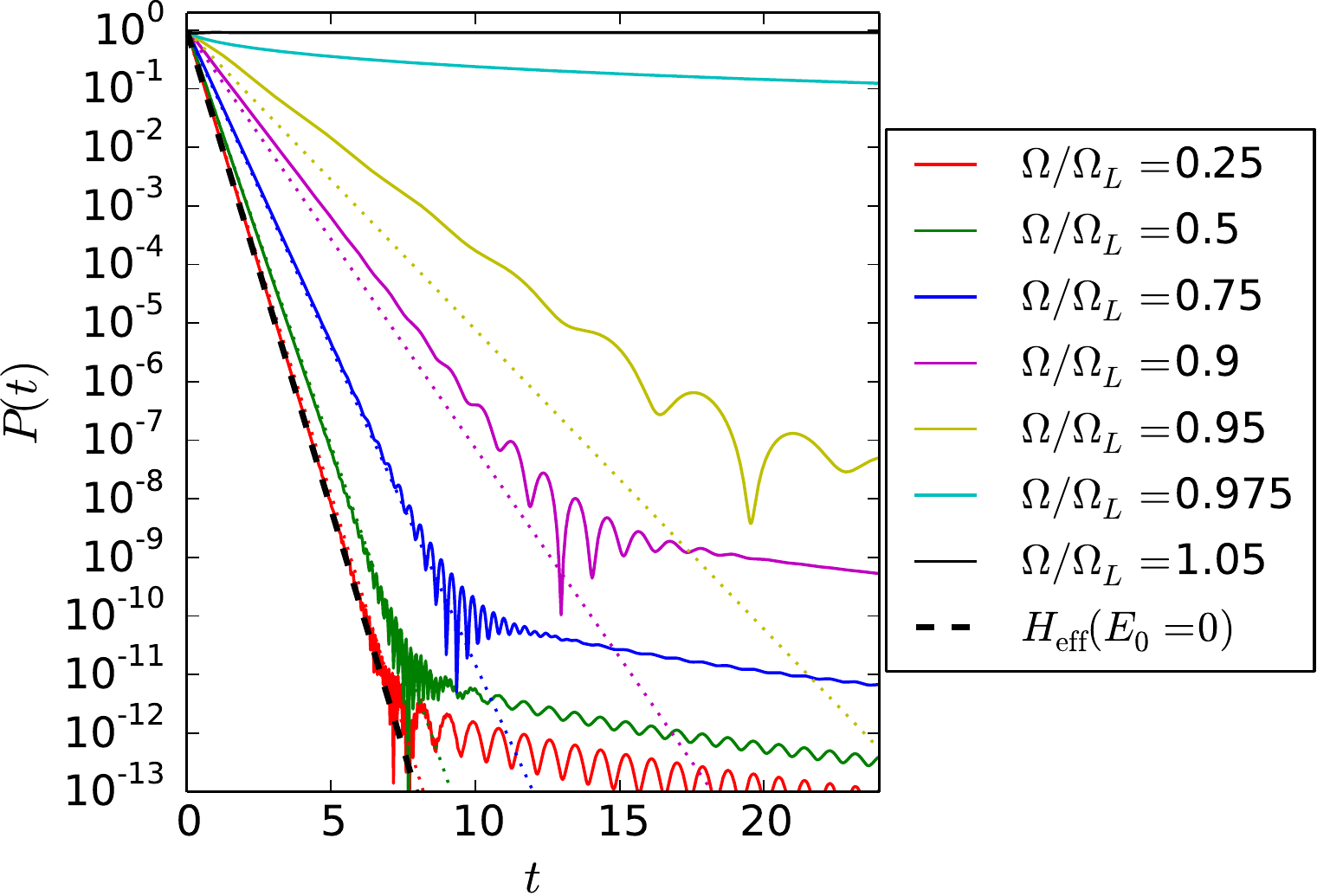}
\caption{(Color online) Survival probability $P(t)$, computed starting from the superradiant state $|S\rangle$ Eq.~\eqref{sr}, is plotted {\it
  vs} time $t$, for different ratios $\Omega/\Omega_L$. The results obtained with the
  full Hermitian model (solid curves) are compared with the exponential decay obtained setting $\Gamma_{\rm sr}(0)=N_R\gamma$ (dashed curve). Dotted lines indicate the exponential decays with decay width given by Eq.~\eqref{eq:Ge0}. 
Data shown refer to the case $N_R = 10$,
$N_L=1000$, $\Omega_L=200$, $\gamma=2$, and different values of
$\Omega$. 
}
\label{fo}
\end{figure}

Here we analyze the situation in which the energy range of the ring can be comparable with the energy band of the lead.
Specifically, we
analyze what happens if $\Omega$ and $\Gamma_{\rm sr}$ are not small
compared to $\Omega_L$, so that the conditions in Eq.~\eqref{eq:condE0} are no longer satisfied.

In Fig.~\ref{fo} we show the survival probability starting from the
state $|S\rangle$ for different values of the ratio
$\Omega/\Omega_L$ and fixed $\Gamma_{\rm sr}$. 
We compare the exact results with the results given by the
non-Hermitian model under the conditions given in
Eq.~\eqref{eq:condE0}.
In the range $\Omega/\Omega_L\le 1/4$ there is a very good agreement 
(compare dashed line in Fig.~\ref{fo}) with numerical results, red solid curve). 
As we increase the ratio $\Omega/\Omega_L$ the exponential decay is
still valid in a significant time range, but the decay width is different.
According to the discussion in Sec.~\ref{sec:eindepH}, since $\Omega$
is not small if compared to $\Omega_L$, we should build our
non-Hermitian Hamiltonian by evaluating $\Delta_{\rm sr}(x)$ and
$\Gamma_{\rm sr}(x)$ at the energy $E_0=2\Omega$,
so that  decay width of the survival probability can be predicted by evaluating $\Gamma_{\rm sr}(x)$ at the energy $2\Omega$ of the initial state:
\begin{equation}\label{eq:Ge0}
\Gamma_{\rm sr}(2\Omega)=N_R\gamma\sqrt{1-\left(\frac{2\Omega}{2\Omega_L}\right)^2}\,,
\end{equation}
which is deduced by the actual form of $\Gamma_{\rm sr}(x)$ given in Eq.~\eqref{eq:GsrApp}.
We plotted such exponential decays as dotted lines in Fig.~\ref{fo}.
Note that the non-Hermitian Hamiltonian derived in the previous Section was obtained by evaluating $\Delta_{\rm sr}(x)$ and
$\Gamma_{\rm sr}(x)$ at the energy $E_0=0$.

When $2\Omega+\Gamma_{\rm sr}(2\Omega)/2 \ge 2\Omega_L$ we have deviations also from the exponential decay obtained by employing the width in Eq.~\eqref{eq:Ge0}. Notice also that as soon as the energy of the initial state is outside the energy band of the lead the decay is strongly suppressed (see data corresponding to $\Omega/\Omega_L=1.05$).
We will see, in the next Section, that the strong suppression of decay when the energy of the initial state lies outside the energy band of the lead will be crucial in understanding the limit of validity of our effective Hamiltonian approximation in presence of disorder.

In our model we have only one state, with energy $2\Omega$, coupled to
the continuum of states in the lead.
Even if the non-Hermitian Hamiltonian model obtained by evaluating $\Delta_{\rm sr}(x)$ and
$\Gamma_{\rm sr}(x)$ at the energy $E_0=2\Omega$ would be valid in a
larger range of parameters, that approximation is not readily
extendable to a situation in presence of disorder, which will be
considered in the following Sections. Indeed, in this case, we do not
have a single level coupled to the lead, but many levels, each with
its own energy. For this reason, we are mainly interested in 
comparing the full dynamics with the
non-Hermitian Hamiltonian model obtained by evaluating $\Delta_{\rm
  sr}(x)$ and $\Gamma_{\rm sr}(x)$ at the energy $E_0=0$, which is
valid when  the dependence on the energy of the initial state can
be neglected and thus it can be easily used also in presence of
disorder.

Note that all of the exponential decays leave place to a power-law
decay above a critical time, which we estimated in the previous
Subsection only in the case $\Omega/\Omega_L\ll 1$. Here we will not
discuss how this transition time is modified as we increase the 
ratio  $\Omega/\Omega_L$, since in this case the deviation from 
non-Hermitian Hamiltonian model obtained by evaluating $\Delta_{\rm
  sr}(x)$ and $\Gamma_{\rm sr}(x)$ at the energy $E_0=0$, occurs for
all times.




\begin{figure}[t]
\centering
\includegraphics[width=8.5cm]{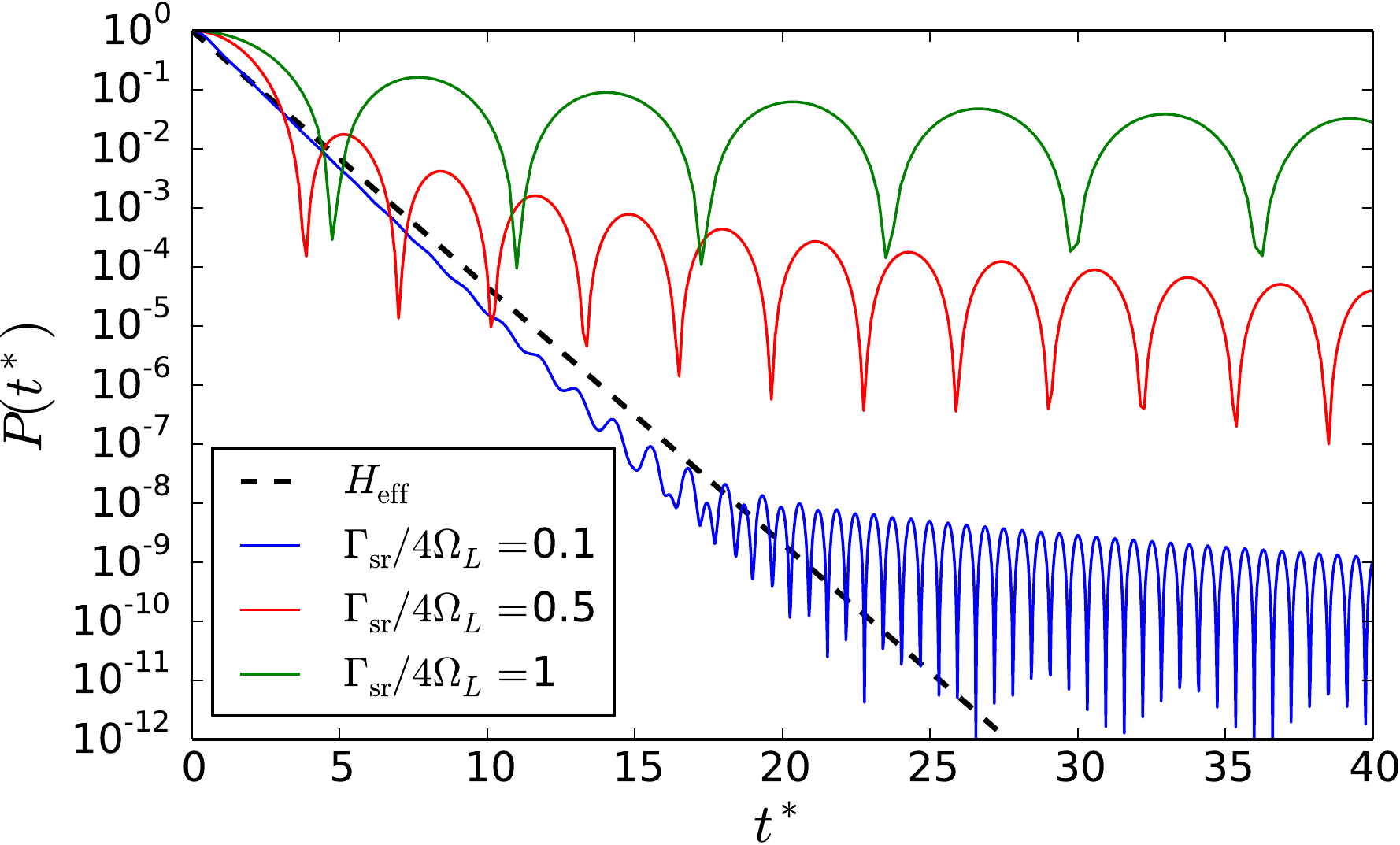}
\caption{(Color online) Survival probability $P(t)$, computed starting from the superradiant state $|S\rangle$ Eq.~\eqref{sr}, is plotted {\it
  vs} the rescaled time $t^*=\Gamma_{\rm sr} t/\hbar=N_R \gamma t/\hbar$ time, for different values of the ratio $\Gamma_{\rm sr}$ over the energy
  bandwidth in the lead $4\Omega_L$. The results obtained with the
  full Hermitian model (solid curves) are compared with the result
  predicted by the non-Hermitian model (dashed curve). 
Data shown refer to the case $N_R = 10$,
$\Omega= 1$,  $N_L=1000$,  $\Omega_L=100$, and different values of
$\Omega_{RL}$. 
}
\label{fG}
\end{figure}

In Fig.~\ref{fG} we show the survival probability starting from the
state $|S\rangle$ for different values of the ratio
$\Gamma_{\rm sr}/4\Omega_L$ and for fixed energy of the initial state in the regime $\Omega/\Omega_L\ll 1$. In this case, deviations from the exponential
decay predicted by the non-Hermitian Hamiltonian start already when
$\Gamma_{\rm sr}/4\Omega_L=0.1$,
thus showing that the agreement between the Hermitian and the
non-Hermitian model is very sensitive to the decay width of the initial
state. The strong oscillations which can be seen in Fig.~\ref{fG} are
due to the fact that, for large $\Gamma_{\rm sr}$, the coupling $\Omega_{RL}$ between the ring and the first lead site is large.

Our results show that, for the non-Hermitian Hamiltonian approach to be effective, the coupling $\Omega_{RL}$ between the ring and the lead does not need to be small with respect to the characteristic energy scale $\Omega$ of the ring, but only with respect to the characteristic energy scale $\Omega_L$ of the lead.

\section{The effects of static disorder}\label{sec:dis}

In this Section we aim at studying the effectiveness of the non-Hermitian Hamiltonian approach in describing the effects of static diagonal
disorder on the transport properties of the system under
consideration. Note that we add disorder only in the ring, leaving the
lead unchanged. Such a disorder is modeled by position-dependent, but
time-independent, fluctuations of the ring site energies, that is we added to the ring Hamiltonian $H_R$, Eq.~\eqref{HR}, the term
\begin{equation}
\label{eq:D}
D=\sum_{r=1}^{N_R}\epsilon_r|r\rangle\langle r|\,,  
\end{equation}
where $\epsilon_r$ are independent random variables uniformly
distributed on $[-W/2,W/2]$, and $W$ represents the disorder strength.

It is well know that, in one-dimensional systems with short-range interactions~\cite{Anderson}, static diagonal disorder induces Anderson
localization: the eigenstates of the system become exponentially
localized. The critical disorder strength
in one-dimensional aggregates for such a localization to occur is given
by 
\begin{equation}\label{eq:fsAnd}
W_{\rm loc}\approx \frac{100}{\sqrt{N}}\,, 
\end{equation}
where $N$ is the system size~\cite{CelGiuBor14}.

To understand the effects of disorder on the excitation transport from the ring into the lead we can make the following considerations: since disorder destroys the perfect symmetry of the
ring, which produces zero decay widths of the subradiant states, it will decrease the decay width of the superradiant state, while it will increase the decay widths of the subradiant states.
Thus, in presence of disorder we do not have only one state
coupled to the lead as in the previous Section, but we have a genuine 
many-level problem.
Note that opening and disorder have competing effects, 
since the opening induces a long-range hopping among the
sites of the ring, as it is clearly seen from the structure of the
full matrix $\mathcal O$ in Eq. \eqref{eq:HeffR} which can be expected
to contrast the localization effects of disorder.

The non-trivial competition between opening and disorder has been analyzed
in Ref.s~\cite{alberto,CelGiuBor14}, within the framework of the
non-Hermitian Hamiltonian approach to open quantum systems. 
It was there
shown that, upon increasing the disorder strength, the decay widths of the
subradiant and superradiant states become the same, and equal to
$\gamma$ for $W > W_{\rm sr}$, where $W_{\rm sr}$ represents the critical disorder
strength above which Superradiance is quenched.

The analysis was performed assuming the
coupling $\gamma$ to the continuum to be  independent of the disorder strength.
Such an assumption is often used in literature and  greatly simplifies the calculations.
On the other side,
one can expect the presence of diagonal disorder to affect the outcome
of the reduction procedure leading to $H_{\rm eff}$, Eq. \eqref{eq:HeffR}. For instance, the coupling to the continuum will in general depend on the disorder strength. 
In order to understand this point, one can consider only one site
coupled to a lead with an energy bandwidth of $4 \Omega_L$. If we
assume the opening strength to the lead to be independent of
disorder, the non-Hermitian Hamiltonian of this system reads $H_{\rm
eff}= E_0 +\epsilon_0 -i\gamma/2$, which implies that the survival
probability decays exponentially as $e^{-\gamma t/\hbar}$ for any value
of the  disorder strength $W$. Clearly, this cannot be true when $W
\gg 4 \Omega_L$, since in that case the probability of the initial
state to be outside the energy band of the lead will be large and the
decay will be consequently suppressed, as discussed in Sec.~\ref{sec:strong}.

Even in presence of disorder, the effective non-Hermitian Hamiltonian can be built following the procedure presented in Sec.~\ref{sec:modelA}, and reads
\begin{equation}
\label{eq:HDx}
H_{\rm eff}(x)=H_R+D+\Delta(x) -\frac{i}{2}Q(x)\,,
\end{equation}
with $Q(x)$ and $\Delta(x)$ given by Eq.~\eqref{eq:QNL} and Eq.~\eqref{eq:DeltaNL}, respectively.
It is clear from that expression that, in the wide-band limit $\Omega_L\to\infty$ (with $\gamma$ fixed), the non-Hermitian Hamiltonian for the disordered ring becomes energy-independent and can be written as
\begin{equation}
H_{\rm eff}=H_R +D-i\frac{\gamma}{2}\mathcal O\,,
\label{eq:HeffRD}
\end{equation}
where $\mathcal O$ is a full matrix with all entries equal to $1$.
Note that this expression coincides with the value for $x=0$ of $H_{\rm eff}$ in Eq.~\eqref{eq:HDx}.

Clearly, the foregoing energy-independent $H_{\rm eff}$ is exact
solely in the infinite-bandwidth limit, while, for any finite
bandwidth in the lead, it will be a good approximation of the true
dynamics only for a disorder strength $W$
sufficiently small if compared to the lead bandwidth, and even in that
case only in a certain time window.
The problem of determining such ranges of validity is very
complicated, since we are dealing with a many-level system and the considerations used in the previous Sections
cannot be used blindly. A discussion of this problem will be given in
the next Subsection.  

The main purpose of this Section is to see whether, for a sufficiently large (but finite) bandwidth in the lead, the important
effects, found in Ref.s~\cite{alberto,CelGiuBor14},
are indeed present in the full Hermitian model
considered in this paper.
The two main findings of Ref.s~\cite{alberto,CelGiuBor14} can be
summarized as follows: (i) Cooperative robustness to disorder. For large enough
opening strength the critical disorder $W_{\rm sr}$ needed to quench Superradiance
increases linearly with the system size. (ii) Subradiant hybrid
regime. In the superradiant regime the response of the superradiant
and subradiant subspaces to disorder is very different. While
superradiant states display robustness to disorder by remaining
extended up to $W_{\rm sr}$, subradiant states show strong signatures
of localization. Indeed, they have hybrid features displaying both an
exponentially localized peak and a uniform delocalized plateau. 
 


\subsection{Comparison between Hermitian and non-Hermitian models in
  presence of disorder}

To assess the effectiveness of the non-Hermitian description, under the assumption that the coupling to the
continuum is independent of disorder, we will first study the survival probability $P(t)$ of
finding the excitation on the ring at time $t$, comparing both the
results given by the Hermitian and the non-Hermitian model in presence
of disorder.

\begin{figure}[t]
\centering
\includegraphics[width=7.2cm]{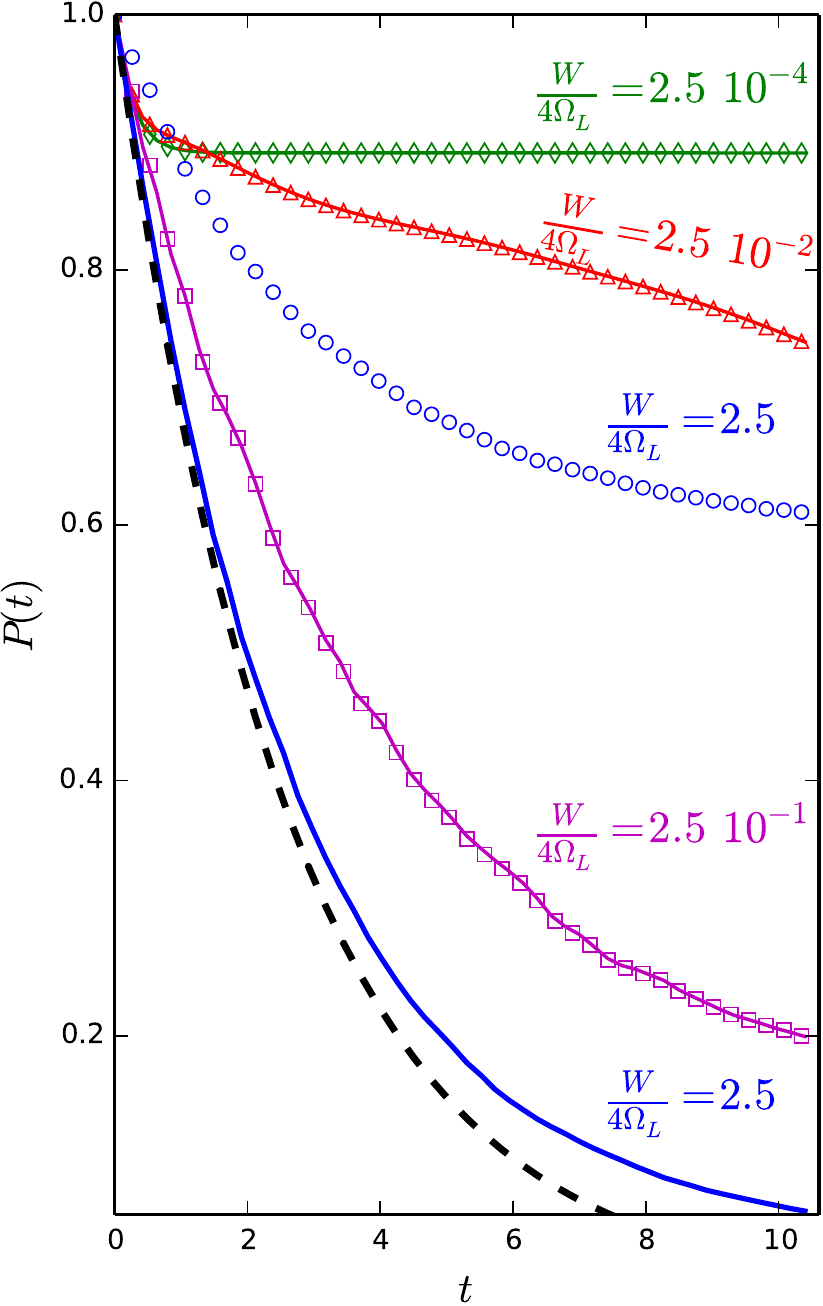}
\caption{(Color online)
Average survival probability $P(t)$, starting from a randomly
generated ring state, computed for different values of the disorder
strength $W$. A perfect agreement between the Hermitian model
(symbols) and the non-Hermitian one (curves) appears, except for the
largest considered value, $W/4\Omega_L=2.5$ (blue), in which case the
prediction of the non-Hermitian model (curve) is remarkably different
from the Hermitian one (circles). As a dashed curve we plotted the
exponential decay $e^{-\gamma t/\hbar}$ corresponding to the
independent sites limit of the non-Hermitian model.
Parameters are $N_R=10$, $\Omega=1$,
$\Omega_{RL}=10$, $\Omega_L=100$, $N_L=250$.
}
\label{fig:rand}
\end{figure} 


We considered a generic initial state generated as a random superposition of the ring eigenstates. 
Since most of the eigenstates are, for sufficiently small disorder, subradiant,
a random initial state will have mainly components on the
subradiant subspace, so that we can expect that disorder will
initially increase the transport efficiency. For very large disorder the
non-Hermitian model predicts an exponential decay of the survival probability $P(t)=e^{-\gamma
t/\hbar}$, while we can expect a much slower decay from the full Hermitian
model.

The average $P(t)$
computed for different values of $W/4\Omega_L$ is shown in Fig.~\ref{fig:rand}.
We observe that, in agreement with the foregoing discussion, when $W/4\Omega_L$ is large, the behavior of the non-Hermitian model departs from that of the Hermitian system for all times. 
Indeed, while the decay of $P(t)$ in the non-Hermitian case is faster and
faster as disorder increases, approaching the limiting decay rate $\gamma/\hbar$, for the Hermitian case the decay has a
non-monotone behavior with the disorder strength, since it increases
for small disorder and it is strongly suppressed for large disorder.

On the other hand, we can see that, for $W/4\Omega_L$ small, our
non-Hermitian model reproduces the Hermitian dynamics in the 
time window shown in the figure.
For any finite bandwidth we expect a departure from
the non-Hermitian description for very small times and very large
times.  Specifically, the time $t_0$ up to which
we have a quadratic decay, can be estimated also in the many-level
case, see Appendix B, where we show that 
$t_0$ is again given by Eq.~(\ref{eq:t0noapp}). On the other side, the time
$t_{S}$ above which we have a departure from the non-Hermitian
Hamiltonian decay is much more difficult to estimate in the many-level case. Indeed, 
the departure from the expoenential decay has
been interpreted in Ref.~\cite{pastawski} as a consequence of the fact
that, for finite bandwidth in the lead, there is a finite return probability
from the lead to the initial state: the transition from
exponential to power-law occurs when the probability to be in the
inital state and the return probability are comparable. In presence of
many levels, the return probability will not only repopulate the
inital state, but also all the other states connected to the
lead. For this reason the estimation of $t_{S}$ in the many-level case
is a delicate issue.

\begin{figure}[t]
\centering
\includegraphics[width=8.5cm]{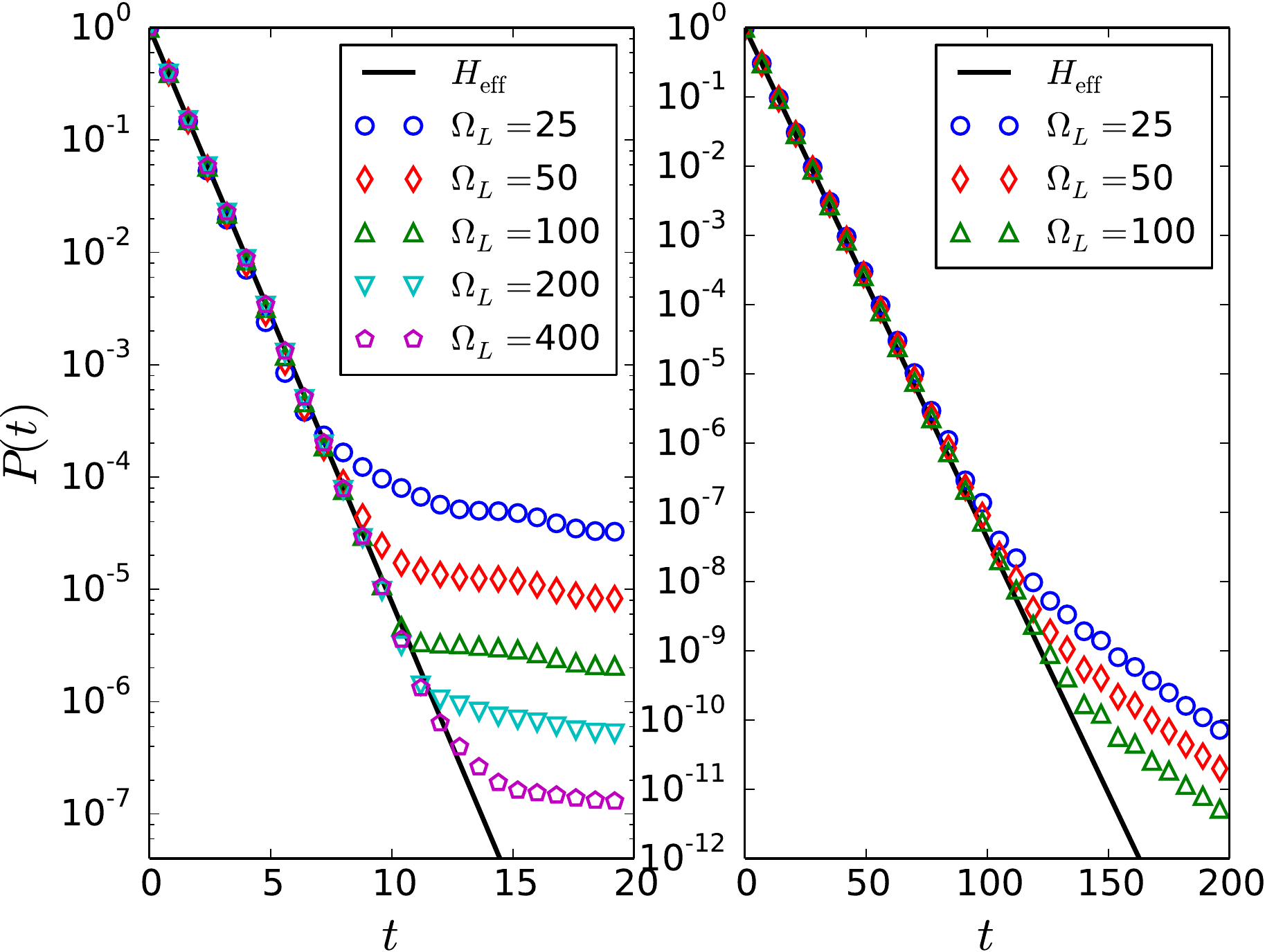}
\caption{(Color online) Survival probability $P(t)$ computed starting from the state with the largest width (left panel) and starting from the state with the second-largest width (right panel) for different values of the coupling $\Omega_L$ within the lead.
A good agreement between the Hermitian model
(symbols) and the non-Hermitian one (curves) is present up to a critical time $t_S$ which increases upon increasing the energy bandwidth ($4\Omega_L$) in the lead.
Parameters are $N_R=4$, $\Omega=1$,
$\gamma=2$, $N_L=4000$, and the disorder strength is given by $W=1$.
}
\label{supsub}
\end{figure}

The rigorous analysis of this problem will be the subject of a future publication,
here we just stress that, as the bandwidth in the lead goes to infinity, we have that $t_0$ goes to zero and $t_S$ grows to infinity. To illustrate this point we analyzed the survival probability $P(t)$ starting from the exact eigenstates of the effective Hamiltonian of Eq.~\eqref{eq:HeffRD}. The dynamics determined by the non-Hermitian model gives an exponential decay of $P(t)$ with a decay width determined by the imaginary part of the complex eigenvalue corresponding to the initial state. In Fig.~\ref{supsub} we compare the non-Hermitian evolution with the Hermitian one obtained from the same initial states as we vary the bandwidth in the lead. In the left panel we show the $P(t)$ computed starting from the state with the largest width (superradiant), while in the right panel we show the $P(t)$ computed starting from the state with the second-largest width (subradiant). In both cases the agreement time $t_S$ between the two models increases as we increase the bandwidth in the lead.

Most importantly, as we decrease the ratio $W/4\Omega_L$ this time window goes to infinity (see Fig.~\ref{supsub}) independently of  the strength of the disorder with respect to the energy scale of the intrinsic system (measured in our case by the ratio $W/4\Omega$).
On the contrary, for any finite bandwidth, as we increase $W$ approaching $\Omega_L$, the evolution given by two models differs for all times. This means that for any given disorder strength the non-Hermitian description will be a good approximation provided that the energy bandwidth of the lead is sufficiently large.

\begin{figure}[t]
\centering
\includegraphics[width=8.5cm]{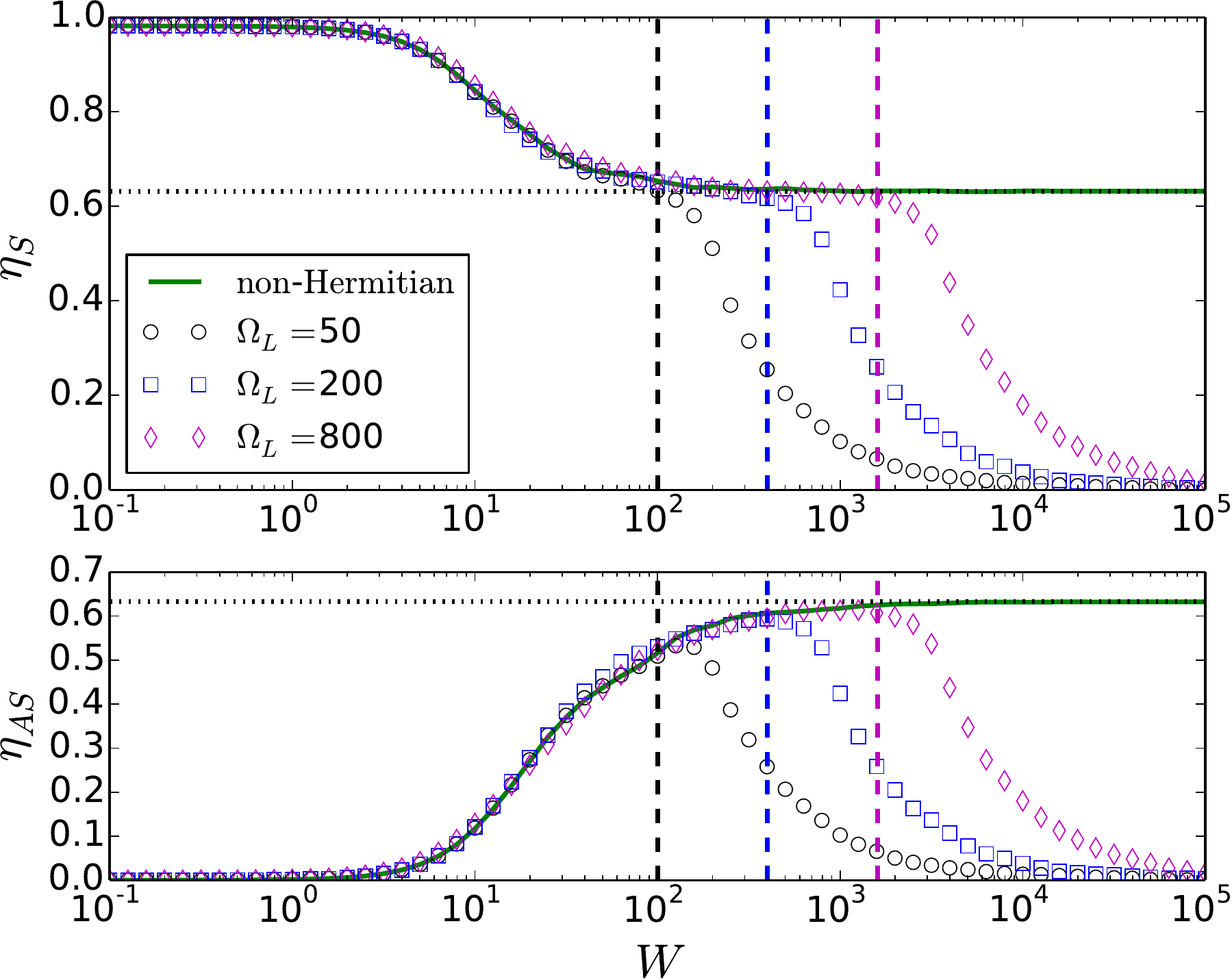}
\caption{(Color online) The efficiency, Eq.~\eqref{eq:eta}, computed starting from the symmetric state of Eq.~\eqref{sr} ($\eta_S$, upper panel) and computed starting from the antisymmetric state of Eq.~\eqref{eq:asym} ($\eta_{AS}$, lower panel) is
  plotted versus the disorder strength $W$. By
  varying $\Omega_L$, we see that the non-Hermitian prediction (full curves) agrees with the Hermitian evolution (symbols) up to a disorder strength (vertical dashed lines) proportional to $\Omega_L$. The dotted horizontal lines
  indicate the non-interacting sites efficiency $1-1/e$.
The parameters used are $N_R=4$, $\Omega=1$, $\gamma=2$, $N_L=\Omega_L$ (to avoid bouncing effects).
}
\label{fw4}
\end{figure} 

To further illustrate this point, we now analyze the agreement between the two models looking at the transport efficiency $\eta(t)$, commonly used in literature, defined as
\begin{equation}
\label{eq:eta}
\eta(t) = 1-P(t)\,.  
\end{equation}
Note that $\eta(t)$ is the probability that the excitation has escaped
into the lead within the time $t$.
In our simulations we set $t=\hbar/\gamma$. If $P(t)$ decays with a rate
$\gamma/\hbar$, corresponding to that of non-interacting decaying sites,
$\eta(\hbar/\gamma)$ assumes the value $1-1/e$. Hence, a value of $\eta(\hbar/\gamma)$ grater than
$1-1/e$ signals a superradiant cooperative decay, while a value of
$\eta(\hbar/\gamma)$ smaller than that threshold signals a subradiant decay.
In what follows we will denote simply by $\eta$ the value $\eta(\hbar/\gamma)$.

In Fig.~\ref{fw4} we show the efficiency $\eta$ varying the coupling $\Omega_L$ in the lead (and accordingly modifying $\Omega_{RL}$ and $N_L$ to keep the decay rate $\gamma$
fixed and remove the bouncing effect). 
The results of the non-Hermitian model are shown as full curves, while
those of the Hermitian model are shown as symbols.
In the upper panel we consider the fully symmetric initial state $|S\rangle$ of Eq.~\eqref{sr}, while in the lower panel we consider the fully antisymmetric state $|AS\rangle$ of Eq.~\eqref{eq:asym}.
For zero disorder the state $|S\rangle$ is superradiant and the state $|AS\rangle$ is subradiant with zero decay width. 

For the non-Hermitian case 
the behavior of $\eta$ is independent
of $\Omega_L$, since we kept $\gamma$ fixed: 
the efficiency of the symmetric state (Fig.~\ref{fw4}, upper panel) decreases with the disorder strength, asymptotically approaching the value $1-1/e$ (dotted line), which would be the efficiency of non-interacting decaying sites, 
while the efficiency of the antisymmetric state (Fig.~\ref{fw4}, lower panel)
increases with the disorder up to the same limiting value.

As for the Hermitian model, it is in perfect agreement with the
non-Hermitian one for small disorder strength, while, for strong
disorder, the efficiency goes to
zero.
This is due to the fact that, when $W>4 \Omega_L$, some of the energy levels in the ring lie outside the energy band in the lead, thus producing a suppression
of decay. Such a suppression is completely neglected in the
non-Hermitian model which is derived by assuming an infinite energy
band in the lead, as explained at the beginning of this Section.
Most importantly, we notice that the agreement between the Hermitian
and the non-Hermitian model increases proportionally to $\Omega_L$, see vertical dashed lines in Fig.~\ref{fw4}.

In the following Subsections we will analyze whether the interesting
effects found in the non-Hermitian model and described at the
beginning of this Section (cooperative robustness and subradiant
hybrid states) can be found also in the Hermitian model for $W\ll\Omega_L$.

\subsection{Cooperative robustness to disorder}\label{sec:coop}
\begin{figure}[t]
\centering
\includegraphics[width=8.5cm]{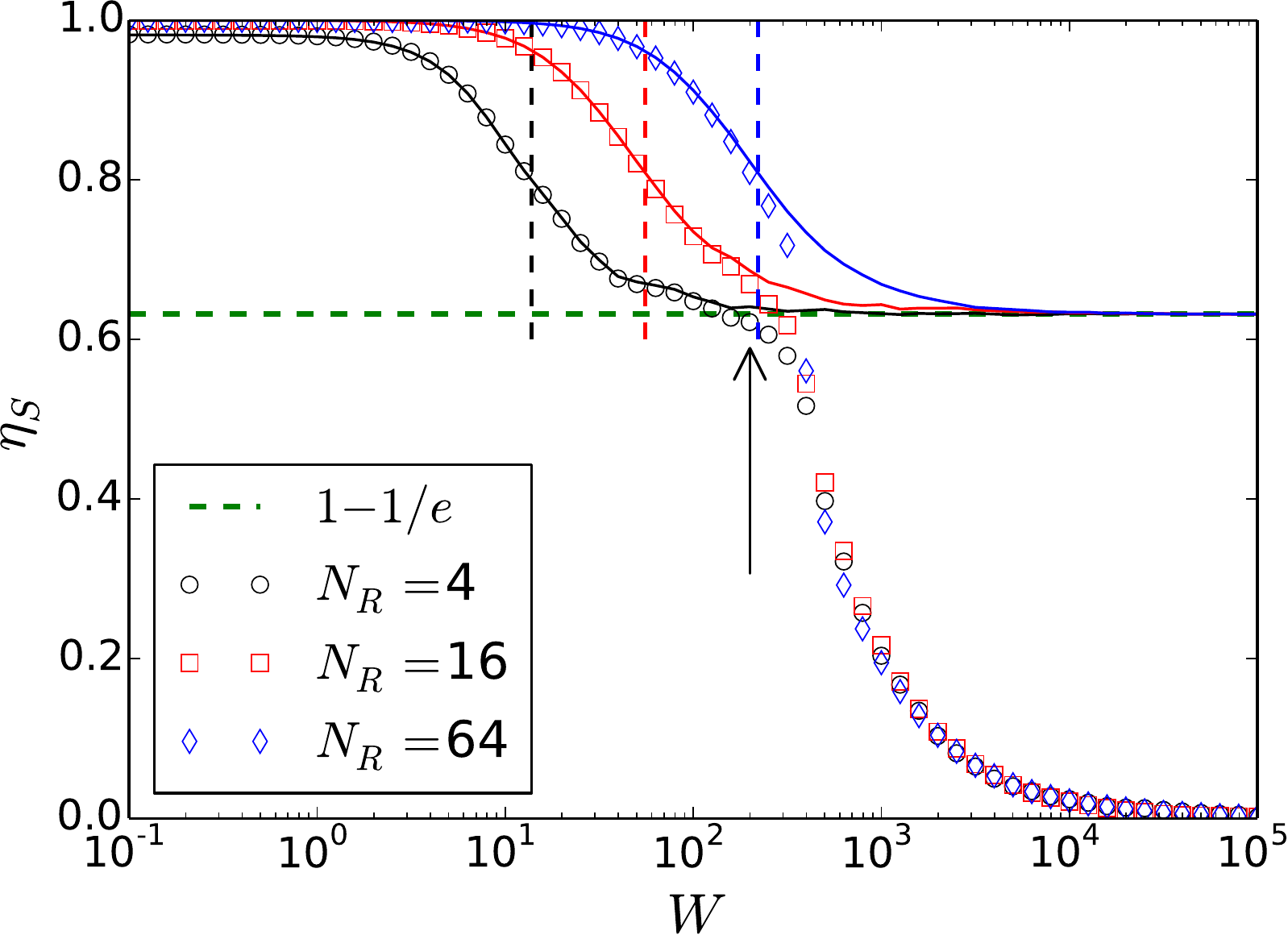}
\caption{(Color online) The efficiency $\eta_S$, Eq.~\eqref{eq:eta}, computed starting from the symmetric state of
  Eq.~\eqref{sr}, is
  plotted versus the disorder strength $W$. The size $N_R$ of the ring has been varied, keeping
  fixed $\Omega=1$, $\Omega_{RL}=10$, $\Omega_L=100$, and
  $N_L=100$. Symbols are obtained with the Hermitian model, while
  solid curves with the non-Hermitian one. The dashed horizontal line
  indicates the non-interacting sites efficiency $1-1/e$, asymptotically
  approached by the non-Hermitian evolution. The vertical dashed lines mark the
  Superradiance transition $W_{\rm sr}$, Eq.~(\ref{eq:wsr}), and the arrow roughly
  indicates the value of disorder up to which the two models agree.
}
\label{fw23}
\end{figure} 

As already mentioned, disorder will quench Superradiance and the
critical disorder $W_{\rm sr}$  at which this occurs has been computed in Ref. \cite[Eq.~(11)]{CelGiuBor14}, for the non-Hermitian model, assuming a
disorder-independent opening strength. 
For the sake of clarity we report below that result:
\begin{equation}
 W_{\rm sr} = \sqrt{\frac{48\Omega^2 (N_R-1)}{\sum_{q=1}^{N_R-1} \frac{1}{\left(\cos \frac{2 \pi q}{N_R}-1\right)^2 +
 \frac{N_R^2 \gamma^2}{16\Omega^2}}}} \,.
\label{Wsr_eq}
\end{equation}
For the parameter range $N_R\gamma\gg 4\Omega$,
Eq.~\eqref{Wsr_eq} reduces to
\begin{equation}
\label{eq:wsr}
W_{\rm sr}=\sqrt{3}N_R\gamma\,.
\end{equation}
 
We stress that the linear growth of $W_{\rm sr}$ with the ring size
$N_R$, Eq.~\eqref{eq:wsr}, is a manifestation of cooperative
robustness to disorder. 

To illustrate this effect we plotted in
Fig.~\ref{fw23} the transport efficiency $\eta$ versus disorder
computed taking as initial state the symmetric state $|S\rangle$, for
different ring sizes $N_R$.
The results for the non-Hermitian model (full curves) are compared
with the results for the Hermitian model (symbols). The agreement
between the two models persists up to a certain value of $W$ indicated by the vertical
arrow in Fig.~\ref{fw23}. The fact that above this value of disorder
the agreement between the two models becomes poor is due to the finite
energy bandwidth in the lead. As it has been explained in the previous
Subsection, the  value of $W$ up to which the to models agree, depends only on $W/4
\Omega_L$, which is kept fixed for the data shown in   Fig.~\ref{fw23}.

Fig.~\ref{fw23} clearly shows that, even in the full Hermitian model, upon increasing the ring size, the
disorder needed to quench the superradiant transport increases. 
That disorder strength is well estimated by $W_{\rm sr}$ given in
Eq.~\eqref{eq:wsr} (see vertical dashed lines in Fig.~\ref{fw23}). 
We also checked that the full expression for $W_{\rm sr}$, Eq.~\eqref{Wsr_eq}, gives a
good estimate of the critical disorder quenching Superradiance in any
parameter range for the Hermitian model provided that $W\ll\Omega_L$.



\subsection{Hybrid subradiant states}\label{sec:hyb}

\begin{figure}[t]
\centering
\includegraphics[width=8.5cm]{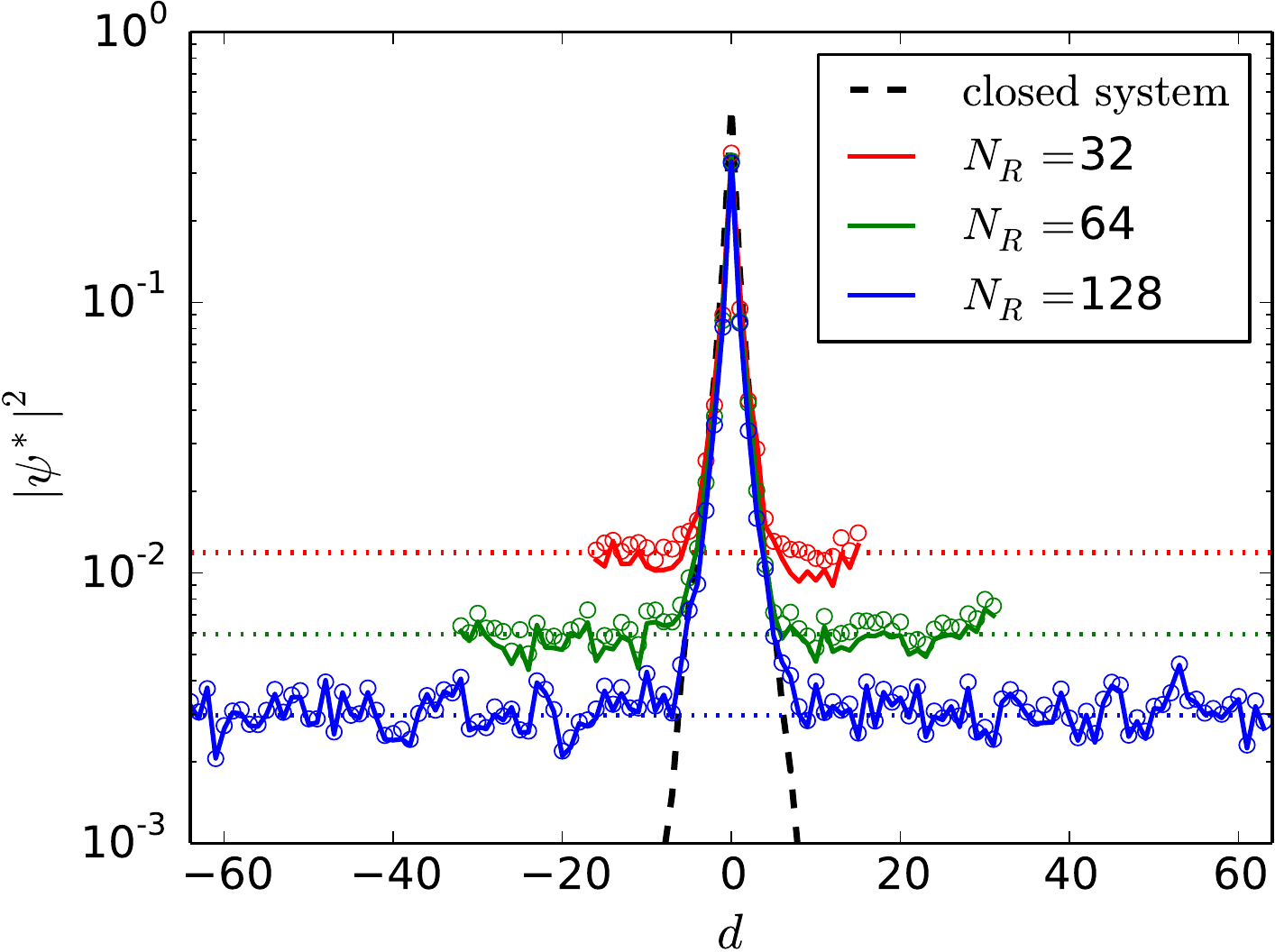}
\caption{(Color online) 
Probability of being on a ring site at distance $d$ from site $1$, obtained by the long-time evolution of an excitation initially localized on site $1$, for different values of the ring size $N_R$. The wave function $\psi^*$ is normalized by setting to $1$ the probability of being on the ring.
We construct the long-time shape of the probability by letting the system evolve until a steady configuration is reached.
We chose the disorder strength $W=10$ in a regime where Anderson localization should be achieved, while Superradiance is not yet destroyed, that is $W_{\rm loc}<W< W_{\rm sr}$. 
Parameters are $\Omega=1$, $\Omega_{RL}=10$, $\Omega_L=100$, $\gamma=2$, and $N_L=100$.
The agreement between the Hermitian model $H+D$ (circles) and the non-Hermitian one $H_{\rm eff}+D$ (solid curves) is very good. The exponential peak on the initially excited site corresponds to the one obtained for a closed ring ($\Omega_{RL}=\gamma=0$), indicated by the black dashed curve.
Dotted horizontal lines mark the values $0.38/N_R$ and have been drawn to highlight the scaling of the plateau with the ring size.
}
\label{fw5}
\end{figure} 

In Ref.~\cite{alberto,CelGiuBor14}  it was
shown that the superradiant state does not localize at the finite-size
Anderson transition, $W_{\rm loc}$ Eq.~\eqref{eq:fsAnd}, but it starts to localize only above the
superradiant transition, $W_{\rm sr}$.  On the other side, 
subradiant states feel the Anderson transition in a way similar to that of the
states of the closed system. Specifically, it was shown that, for
$W_{\rm loc}<W<W_{\rm sr}$, subradiant
states display a hybrid nature, with an exponentially localized
peak and an extended plateau. The persistence of signatures of Anderson localization in the
subradiant regime is somehow surprising: since in this regime the
opening is large, one could expect that the long-range
coupling induced by the opening would destroy localization.
This regime was named subradiant hybrid regime in Ref.~\cite{alberto}.

To show that this regime is present also in the Hermitian model,
we cannot follow the same procedure that was followed in
Ref.~\cite{alberto,CelGiuBor14}, where the structure of the eigenstates
of the effective Hamiltonian was analyzed. On the other side we can
analyze the long-term dynamics of a state initially localized on a
single site of the ring. This state has a small overlap with the
superradiant state. That component will decay fast, and the dynamics
will bring the system in the subradiant subspace with a much
slower decay. 
Thus we can expect that the hybrid structure of the subradiant states will reveal itself in the long-time form of the wave-function.

In order to show this point,
in Fig.~\ref{fw5} we plot the probability of being on the ring site
$r$, obtained by the long-time evolution of an
excitation initially localized on site $1$. We chose the disorder strength $W$
in a regime where Anderson localization should be achieved, while
Superradiance is not yet destroyed, that is $W_{\rm loc}<W< W_{\rm sr}$. Of course, we chose a value of $\Omega_L$ for which the agreement between the Hermitian and the non-Hermitian model is good in the relevant disorder range.
The probability plotted in Fig.~\ref{fw5} is normalized by setting to $1$ the probability of being on the ring. 

In the localized regime and in absence of the coupling with the lead
the diffusion of an excitation initially placed on one
site would be suppressed, resulting in a long-time probability
distribution exponentially localized on the initial site (see dashed
curve in Fig.~\ref{fw5}).

On the other side, in the full model we obtain a hybrid state, characterized by an
exponential peak on the initial site and a fully extended plateau on
the other sites. The
important features of this hybrid structure are: (i) the exponential
peak coincides with the one obtained 
in a closed ring (for which $\Omega_{RL}=\gamma=0$); (ii) the probability on the extended
plateau decreases as $1/N_R$ as we increase the ring size. Again we
observe that the non-Hermitian model (solid curves) is in very good
agreement with the Hermitian one (circles), thus proving that the
presence of hybrid subradiant states, described in~\cite{alberto}, is
a genuine feature of the full Hermitian model from which $H_{\rm eff}$
is deduced. 
Importantly, in the limit $N_R\to\infty$, the subradiant states become
fully localized. For a more detailed discussion
of the origin of this regime see Ref. \cite{alberto}.

\section{Transport in a generic network}\label{sec:random}

\begin{figure}[t]
\centering
\includegraphics[width=5.cm]{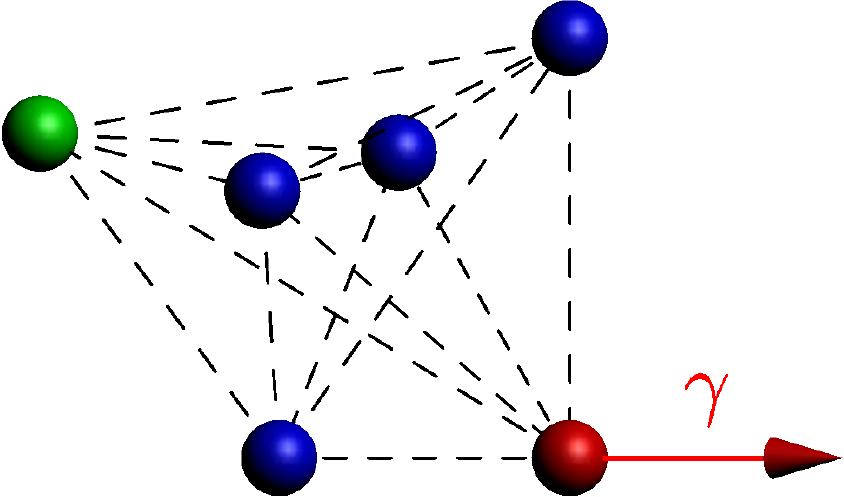}

\vspace{0.8cm}

\includegraphics[width=8.5cm]{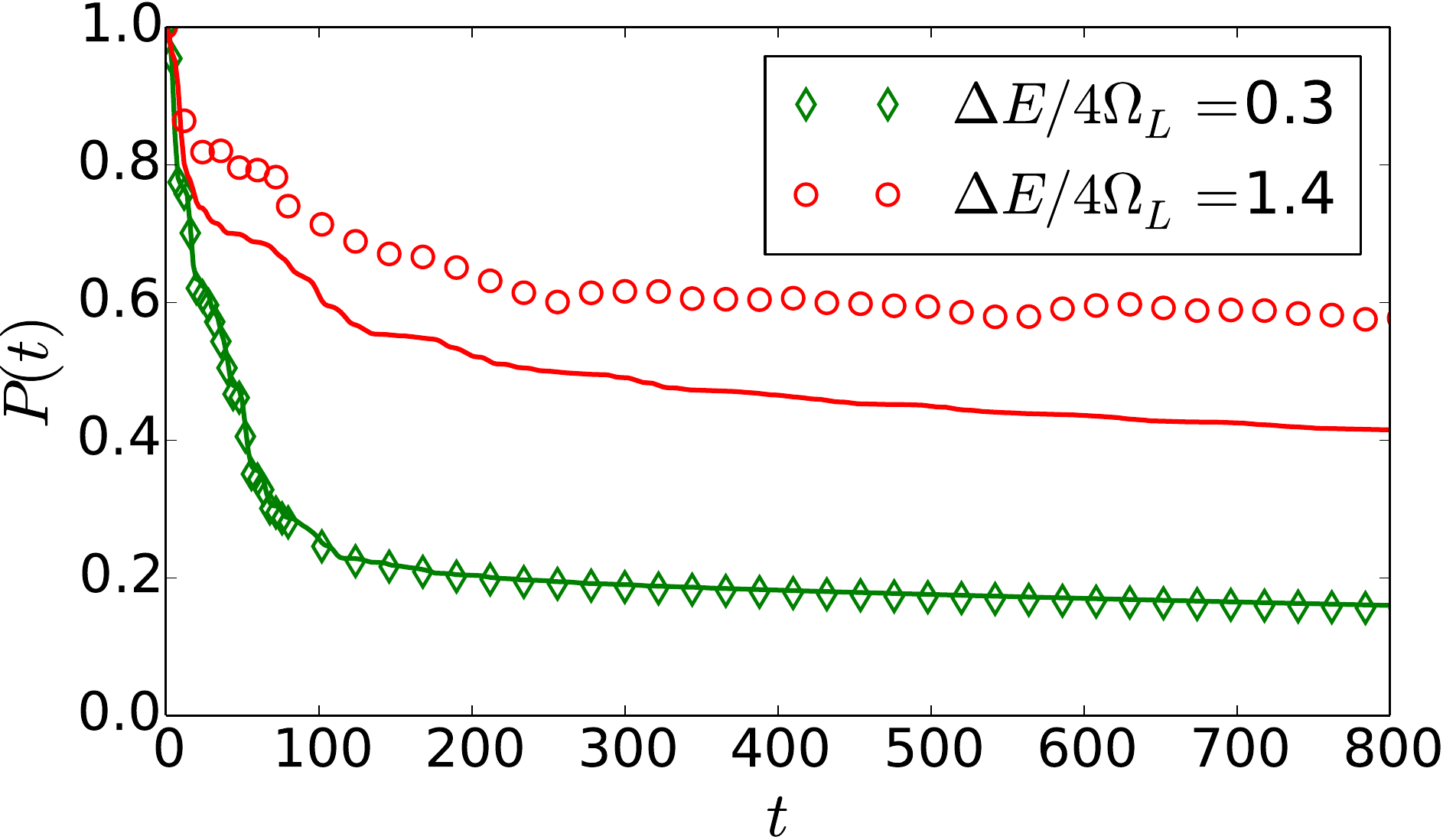}
\caption{(Color online) In this figure we consider a fully-connected disordered network, depicted in the upper panel and described in Sec.~\ref{sec:random}. One of the sites (red bottom-right site) is coupled to an external one-dimensional lead with $N_L=4000$. In the lower panel we compare the results of the Hermitian model (symbols) with those of the non-Hermitian one (curves). 
We analyze the probability $P(t)$ of being in the network {\it vs} time, computed for an initial state localized on one site (green upper-left site, upper panel) which is not directly connected to the lead. Data shown refer to a single realization of the disordered network and they have been obtained by changing the ratio $\Delta E/4\Omega_L$ between the energy range of the disordered network and the energy band in the lead (see legend).
Note that we kept the value $\gamma=2$ constant and we chose $W=1$.
}
\label{network}
\end{figure}

In order to discuss the range of applicability of the results
discussed so far, we consider here a generic example of network with a
sink represented by an external one-dimensional lead. From our
previous results we expect that the energy-independent non-Hermitian Hamiltonian approach will be valid under the condition
that the energy band in the lead is much larger than the energy range of the network.

We consider a fully-connected disordered network, depicted in the upper panel of Fig.~\ref{network}. The system is described by a tight-binding Hamiltonian with site energies chosen randomly in the interval $[-W/2,W/2]$. Moreover, each site is coupled to all of the other sites with a tunneling coupling $\Omega$ randomly distributed in the interval $[-1,1]$. 
One of the sites is coupled to an external one-dimensional lead with coupling $\Omega_{RL}$. The external lead is an ordered chain of sites connected with tunneling coupling $\Omega_L$. 

The non-Hermitian Hamiltonian for this model can be obtained, following Sec.~\ref{sec:model}, by adding to the energy of the site connected to the lead the imaginary term $-i\gamma/2$, with $\gamma$ given in Eq.~\eqref{eq:gamma}.
The diameter in the complex plane of the eigenvalues of the non-Hermitian Hamiltonian defines the energy range $\Delta E$ of the disordered network, while the energy bandwidth in the lead is given by $4\Omega_L$. We thus expect the non-Hermitian approximation to be effective when $\Delta E/4\Omega_L$ is sufficiently small. 

To show this point 
we analyze the probability of being in the network {\it vs} time, computed for an initial state localized on one site which is not directly connected to the lead. 
The typical result is shown in Fig.~\ref{network}, lower panel: when
$\Delta E/4\Omega_L<1$ the Hermitian and non-Hermitian models agree
(compare diamonds with green curve) in a large time window; on the other side, when $\Delta E/4\Omega_L>1$, the non-Hermitian model looses its validity (compare circles with red curve).

\section{Conclusions}\label{sec:conclusions}

We analyze the problem of describing the transport properties of quantum networks coupled to external environments acting as sinks, in the sense that they absorb the excitation from the network in an irreversible way.
To this end, we analyze a paradigmatic model for quantum transport and decay. Our tight-binding model consists of a network of sites arranged in a
ring and connected to a central lead. We derive an
energy-independent non-Hermitian model which greatly simplifies the analysis of its transport properties. 
This non-Hermitian model retains only the degrees of freedom of the ring, summarizing the coupling with the infinite degrees of freedom of the lead into non-Hermitian opening terms, which induce a decay of the probability to be on the ring. Such non-Hermitian terms can be obtained from the same quantities which are used in the Fermi Golden Rule: the transition amplitudes from the discrete states of the quantum network to the continuum of states in the external sinks, and the density of states in the sinks.

Such a kind of non-Hermitian models are widely
used in literature, but the problem of their validity is often
overlooked. Here, by comparing the results of the full
Hermitian model with those given by the non-Hermitian one, we demonstrate that the
energy-independent non-Hermitian Hamiltonian approach is valid in the
regime of large energy band in the lead.
Under that condition, we show 
that the interesting effects usually
described with the non-Hermitian model, such as Super and Subradiance
in transport, are present also in the full Hermitian model. 

We also consider the decay from the ring in presence of static
disorder. 
We discuss the validity of the assumption that the
opening strength to the continuum is independent of disorder, which is often used in literature since it greatly simplifies the problem.
We show that the non-Hermitian Hamiltonian, with
opening terms independent of disorder, is able to describe the decay in the
full Hermitian model for a range of disorder for which the energy range
in the ring is much smaller than the energy band in the lead. 
In this regime, we were able to confirm the existence of the interesting effects predicted within the non-Hermitian Hamiltonian approach also in the full Hermitian model.
Indeed, superradiant states are cooperatively
robust to disorder, while subradiant states show a different
behavior, displaying a hybrid nature, due to the interplay of
disorder and opening.

Our results have a wide range of applicability: if we consider a
generic quantum network of sites coupled to an external lead,
the energy-independent non-Hermitian Hamiltonian approach
is valid under the condition
that the energy band in the lead is much larger than the energy range of the network. 
Specifically, in this limit it can give an accurate description of any observable of the network.
In the case of generic external environments acting as sinks, 
the same approach is effective
when the transition amplitudes from the network states to the sink
states are smooth and slowly varying functions of the energy, in the range determined by the eigenvalues of the disordered network. Moreover, we want to stress that, for this approach to be valid, the coupling between the system and the environment does not need to be small with respect to the characteristic energy scale of the system, but only with respect to the characteristic energy scale of the external environment.

\section*{Acknowledgments}

We would like to thank F. Borgonovi for his valuable comments on earlier drafts of the present paper.
We also acknowledge useful discussions with A. Biella,
R. Fazio, L. Kaplan, F. Izrailev, 
S. Pascazio, and H. M. Pastawski.

\appendix
\section{}

In order to estimate the effect of the finite bandwidth on the decay, we consider a different approximation of the time-evolution operator of Eq.~\eqref{eq:Usr}, slightly more refined than the one leading to $H_{\rm eff}$ and Eq.~\eqref{eq:UteffR}.
We assume the bandwidth to be finite, but large enough to justify the
following approximation: we consider the transition amplitude
$A_r(E)$ and the density of states $\rho(E)$, see Eq.~\eqref{eq:AA}, to be constant in the
finite energy band $[-2\Omega_L,2\Omega_L]$. Specifically, we assume
that $A_r(E) (A_{r'}(E))^* \rho(E)=\gamma/ 2 \pi$ inside the energy
band of the lead and zero outside.  
We can now substitute in Eq.~\eqref{eq:Usr} the limiting values given by
Eq.~\eqref{eq:limDG}, 
that corresponds to choosing
\begin{equation}\label{eq:Gsr}
\Gamma_{\rm sr}(x)=\left\{
\begin{array}{cl}
\gamma N_R & \quad\text{for }x\in[-2\Omega_L,2\Omega_L]\,,\\
&\\
0 & \quad\text{otherwise,}
\end{array}\right.
\end{equation}
Moreover we also assume $\Delta_{\rm sr}(x)=0$. The evolution operator of Eq.~\eqref{eq:Usr} becomes then
\begin{equation}
\label{eq:Uxindep}
\mathcal U_S(t,0)\approx\frac{1}{2\pi Z}\int_{-2\Omega_L}^{2\Omega_L}
\frac{e^{-\frac{i}{\hbar}xt}\gamma N_R}{[x-2\Omega]^2+\frac{1}{4}\gamma^2N_R^2}\,dx\,,
\end{equation}
that is the Fourier transform of a truncated Lorentzian, suitably normalized by means of the factor $Z$ to ensure that $\mathcal U_S(0,0)=1$.
The evolution will be well approximated by
an exponential only for intermediate times and we will have deviations both for
small and large times, due to the truncation at the edges of
the energy band of the lead.

Some remarks on the accuracy of the approximation leading to Eq.~\eqref{eq:Uxindep} are in order.
Given the choice of $\Gamma_{\rm sr}(x)$ in Eq.~\eqref{eq:Gsr}, it is possible to explicitly compute the energy shift
\begin{equation}
\label{eq:Dsr}
\Delta_{\rm sr}(x)=\gamma \ln\frac{|x+2\Omega_L|}{|x-2\Omega_L|}\,.
\end{equation}
The latter function is odd, with derivative
\[
\Delta'(0)=\frac{\gamma}{2\pi\Omega_L}\,,
\]
and slowly divergent as $x$ approaches the edges of the band.
We then see that, by setting $\Delta_{\rm sr}(x)=0$, we obtain an integrand in Eq.~\eqref{eq:Uxindep} significantly distorted if compared to the exact one (Eq.~\eqref{eq:Usr}) only in a neighborhood of the edges of the band.
To minimize the effects of such a distortion, it is then crucial that
the maximum point of the exact integrand function lies far enough from
the edges of the energy band of the lead. Moreover we need the decay
width of the superradiant state to be much smaller than the energy band
in the lead. Since the position of the maximum point is determined (to leading order) by
the average energy of the Superradiant state
$\bra{S}H\ket{S}=2\Omega$, we obtain the conditions:
\begin{equation}
\label{eq:condE0app}
\Omega\ll\Omega_L\,,  \qquad N_R \gamma \ll 4 \Omega_L\,.
\end{equation}
These conditions are necessary for the approximation leading to Eq.~\eqref{eq:Uxindep} to be accurate.

Starting from the evolution operator 
obtained in Eq.~\eqref{eq:Uxindep}, we can now give an estimate of the times $t_0$ and $t_S$ at which the decay of the survival probability $P(t)$, computed with the Hermitian evolution of $|S\rangle$, changes from the quadratic behavior to the exponential decay predicted by the non-Hermitian model ($t_0$) and from the exponential decay to a power-law decay ($t_S$).

The evolution operator $U_S(t,0)$ of Eq.~\eqref{eq:Uxindep} is given by the Fourier transform of a Lorentzian function multiplied by a rectangular function with support on $[-2\Omega_L,2\Omega_L]$. Recalling that the Fourier transform of that rectangular function is given by 
\[
\frac{\sin\left(\frac{2\Omega_L}{\hbar}t\right)}{\pi t}
\]
and that the Fourier transform of a product is the convolution of the Fourier transforms,
Eq.~\eqref{eq:Uxindep} becomes
\begin{equation}
\label{eq:Uconv}
\mathcal U_S(t,0)=\int_{-\infty}^{+\infty}\frac{\sin\left(\frac{2\Omega_L}{\hbar}\tau\right)}{Z\pi\tau}e^{-\frac{2\Omega i}{\hbar}(t-\tau)}e^{-\frac{\gamma N_\mathit{R}}{2\hbar}|t-\tau|}\,d\tau\,.
\end{equation}
Now, since
\[
\lim_{\omega\to\infty}\frac{\sin(\omega\tau)}{\pi\tau}=\delta(\tau)\,
\]
for any $\omega$ in the sense of distributions, if we consider in Eq.~\eqref{eq:Uconv} the wide-band limit $\Omega_L\to\infty$, we immediately recover the evolution given by Eq.~\eqref{eq:UteffR} for any time $t$. On the other hand, the effects of a finite bandwidth strongly modify the decay at both small and large times.

Under the assumption of Eq.~\eqref{eq:condE0app}, we can set $\Omega\approx 0$ and neglect the oscillating term $e^{-\frac{2\Omega i}{\hbar}(t-\tau)}$, so that $\mathcal U_S(t,0)$ reduces to the convolution of the exponential decay $e^{-\frac{\gamma N_\mathit{R}}{2\hbar}|t|}$ with the kernel
\begin{equation}\label{eq:kernel}
K(\tau)=\frac{\sin\left(\frac{2\Omega_L}{\hbar}\tau\right)}{Z\pi\tau}\,.
\end{equation}
The normalization factor $Z$, needed to compensate the approximation $\Delta(x)\approx 0$ already introduced in Eq.~\eqref{eq:Uxindep}, can be easily found by applying the normalization condition $\mathcal U_S(0,0)=1$.

The fact that the small-time decay is quadratic can be easily seen by considering the parity of $e^{-\frac{\gamma N_\mathit{R}}{2\hbar}|t|}$ and $K(\tau)$: since they are both even functions, their derivatives are odd and
\[
\int_{-\infty}^{+\infty}\frac{\sin\left(\frac{2\Omega_L}{\hbar}\tau\right)}{Z\pi\tau}\left.\frac{d}{dt}\left(e^{-\frac{\gamma N_\mathit{R}}{2\hbar}|t-\tau|}\right)\right\vert_{t=0}\,d\tau=0\,.
\]
Consequently, the derivative of $\mathcal U_S(t,0)$ vanishes for $t=0$ and the decay is quadratic. This is true for any finite value of $\Omega_L$, but we see a sharp transition to a linear short-time decay in the limit $\Omega_L\to\infty$.

\begin{figure}[t]
\centering
\includegraphics[width=8.5cm]{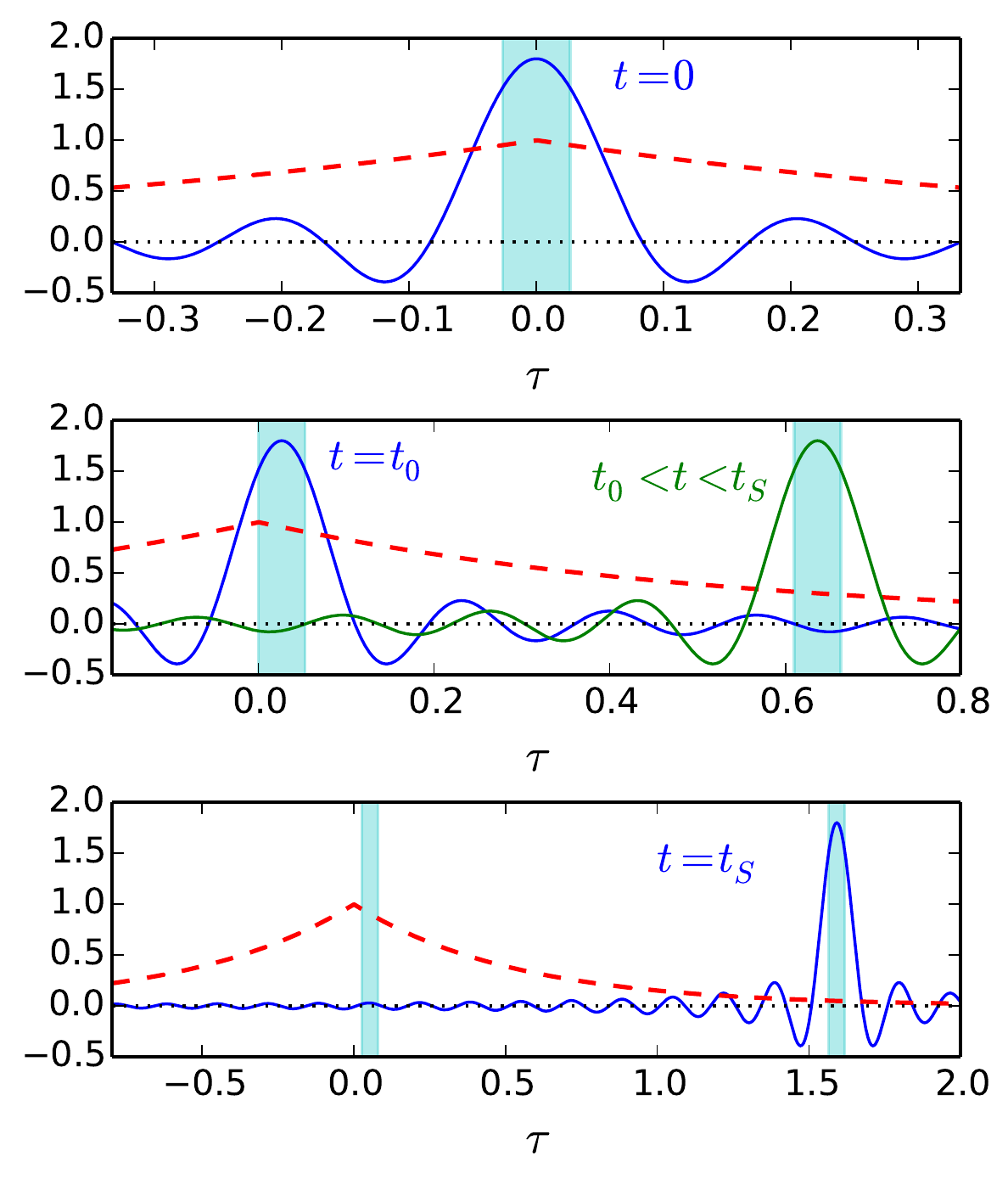}
\caption{(Color online) To illustrate the argument presented in this Section we plotted the exponential function $e^{-\frac{\gamma N_\mathit{R}}{2\hbar}|\tau|}$ (dashed red curve) and the kernel $K(\tau-t)$ of Eq.~\eqref{eq:kernel} (solid curves) for different times $t$. The shaded regions are those providing the dominant contribution to the convolution product of Eq.~\eqref{eq:Uconv} at each time.}
\label{fig:convoluzioni}
\end{figure}

An intuitive explanation of that transition can be given with the aid of Fig.~\ref{fig:convoluzioni}. In the first panel we plotted $e^{-\frac{\gamma N_\mathit{R}}{2\hbar}|\tau|}$ and $K(\tau-t)$ for $t=0$. The evolution at each time $t$ is given by the integral of the product of the exponential and the kernel $K$ and the dominant contribution for $t=0$ comes from the shaded region in Fig.~\ref{fig:convoluzioni} first panel.
The amplitude of this region is twice the inverse of the oscillation frequency $2\Omega_L/\hbar$. 
As we increase $t$ of a small amount $dt$, since the shaded region lies on both sides of the peak of the exponential function, the variation in the integral is of order $O(dt^2)$, producing a quadratic decay. This is no longer true in the limit $\Omega_L\to\infty$, since $K$ tends to a Dirac function whose support can lie only on one side of the peak, so that the variation of the convolution integral is of order $O(dt)$, entailing a linear small-time decay.

From analogous considerations we can estimate $t_0$ as the time at which the relevant region (shaded region in Fig.~\ref{fig:convoluzioni}, second panel) lies only on one side of the peak of the exponential. Since the peak of the kernel $K(t-\tau)$ is at $\tau=t$, we have
\begin{equation}
\label{eq:t0}
t_0=\frac{\hbar}{2\Omega_L}\,.  
\end{equation}
We can then see that, for $t>t_0$ the decay is exponential, since the main contribution to the convolution integral (see shaded regions in Fig.~\ref{fig:convoluzioni}, second panel) is proportional to
\[
e^{-\frac{\gamma N_\mathit{R}}{2\hbar}|t|}\,.
\]

For even larger times, together with the previously described exponential term (right shaded region in Fig.~\ref{fig:convoluzioni}, third panel), a second term contributes to the convolution integral (left shaded region in Fig.~\ref{fig:convoluzioni}, third panel). The first involves the central part of the kernel $K$ and the tail of the exponential function, while the second involves the tail of the kernel $K$ and the central part of the exponential function. The first contribution is again proportional to $e^{-\frac{\gamma N_\mathit{R}}{2\hbar}|t|}$ and the second one to $\hbar/(2\Omega_L t)$.
When the term involving the tail of $K$ is dominant, we have a power-law decay.
Hence, we can estimate the transition time $t_S$ as the time at which the two contributions are comparable by setting
\[
e^{-\frac{\gamma N_\mathit{R}}{2\hbar}t_S}\approx \frac{\hbar}{2\Omega_L t_S}\,,
\]
which leads to the equation
\[
\frac{\gamma N_R}{2\hbar}t_S=\ln\frac{4\Omega_L}{\gamma N_R}+\ln\frac{\gamma N_R}{2\hbar}t_S
\]
and, neglecting the last logarithmic term, to the estimate
\begin{equation}
\label{eq:tSapp}
t_S\propto  \frac{2\hbar}{\gamma N_R}\ln\frac{4\Omega_L}{\gamma N_R}\,.
\end{equation}

The exponent of the power-law decay, being determined by the long-time behavior of the convolution kernel $K$, is strongly dependent on how the Lorentzian density of Eq.~\eqref{eq:Uxindep} is deformed to be zero outside the energy band of the lead.
Indeed, the sharp truncation considered above, given by the definition of $\Gamma_{\rm sr}$ in Eq.~\eqref{eq:Gsr}, corresponds to multiplying the Lorentzian with a rectangular function, that produces a $1/t$ decay due to the form of the kernel $K$ of Eq.~\eqref{eq:kernel}. 

Nevertheless, we can easily understand the effect of a different deformation: if we multiply the Lorentzian function in Eq.~\eqref{eq:Uxindep} by a compactly supported function which goes to zero as $(x-E_{\rm edge})^{p}$ in proximity of the edges of the energy band, by a well-known result in Fourier analysis~\cite{pascazio,bochner}, we will obtain a convolution kernel $K$ which decays as $1/t^{p+1}$ for large times.
Such a modification does not affect any of the foregoing results, but produces a long-time decay of the probability amplitude proportional to $1/t^{p+1}$. Consequently, the survival probability $P(t)$ will decay as $1/t^{2(p+1)}$.

If we consider now the detailed structure of $\Gamma_{\rm sr}$ in the finite-bandwidth case, Eq.~\eqref{eq:GsrApp}, we see that it goes to zero in proximity of the edges of the energy band  with exponent $p=1/2$. This implies a decay $1/t^{3/2}$ of the convolution kernel and the decay $1/t^3$ of the survival probability $P(t)$, which was indeed found in the numerical results shown in Fig.~\ref{f22}.



\section{}

To determine the behavior for very short times we will follow now a
different approach, more heuristic than the one used in Appendix A.
If we consider an initial state on the ring, then it ``becomes aware'' of the presence of the lead only after some time. In particular, we can expect the initial dynamics to be determined by the interaction of the ring with the first site of the lead. If we had $\Omega_L=0$, the fully symmetric ring state $\ket{S}$ would be only coupled to the first lead site, and its dynamics would be determined by the $2\times 2$ Hamiltonian
\[
\begin{pmatrix}
2\Omega & \sqrt{N_R}\Omega_{RL} \\
\sqrt{N_R}\Omega_{RL} & 0
\end{pmatrix}\,,
\]
which has eigenvalues 
\[
\lambda_{1,2}=\Omega\pm\sqrt{\Omega^2+N_R\Omega_{RL}^2}\,. 
\]
Consequently, 
we obtain the following estimate for the short-time decay of the survival probability of the superradiant state:
\begin{equation}\label{eq:sht}
P(t)\approx 1-\frac{N_R\Omega_{RL}^2}{\hbar^2}t^2\,.
\end{equation}

According to the foregoing argument, the time $t_0$ up to which Eq.~\eqref{eq:sht} can be a good approximation of the dynamics should decrease upon increasing the coupling $\Omega_L$ within the lead. Indeed, the value
$t_0=\hbar/2\Omega_L$,
presented in Eq.~\eqref{eq:t0}, gives a good estimate of this threshold.
Clearly, the non-Hermitian model cannot reproduce the true dynamics of the system up to $t_0$, since that model is obtained considering the effect of a lead with an infinite coupling $\Omega_L$ in the lead.

Let us now consider the case of a disordered ring described by the Hamiltonian $H+D$, Eq.s~(\ref{eq:HRL}, \ref{eq:D}). Also in this case, for short times the true evolution will be different from the evolution given by the non-Hermitian model. Indeed, the short-time dynamics
is well approximated by the evolution under $\widetilde{H}+D$, where $\widetilde{H}$ describes the subsystem formed by the ring and the first lead site. We can estimate with the same $t_0$ given above the time up to which the system does not feel the presence of the other lead sites and, consequently, the non-Hermitian model is not applicable.

\end{document}